\documentclass[]{aa}
\usepackage{color }
\usepackage{natbib}
\usepackage{amsmath}
\usepackage{amssymb}
\usepackage{graphicx}
\graphicspath{{pictures/}}
\DeclareGraphicsExtensions{.png}
\newcommand{\bl}[1]{\mbox{\boldmath$ #1 $}}

\begin{document}

\title{Accretion bursts in magnetized gas-dust protoplanetary disks}

\author{Eduard I. Vorobyov\inst{1,2}, Sergey Khaibrakhmanov\inst{3,4}, Shantanu Basu\inst{5}, and Marc Audard\inst{6}}
\institute{ 
University of Vienna, Department of Astrophysics, Vienna, 1180, Austria \\
\email{eduard.vorobiev@univie.ac.at} 
\and
Research Institute of Physics, Southern Federal University, Rostov-on-Don, 344090 Russia
\and 
Ural Federal University, 51 Lenin Str., 620051 Ekaterinburg, Russia
\and 
Chelyabinsk State University, Theoretical Physics Department, 454001 Chelyabinsk, Russia
\and
University of Western Ontario, Department of Physics and Astronomy, London, Ontario, N6A 3K7, Canada
\and
University of Geneva, Department of Astronomy, Chemin d’Ecogia 16, 1290 Versoix, Switzerland
}

\abstract
{}
{Accretion bursts triggered by the magnetorotational instability (MRI) in the innermost disk regions were studied for protoplanetary gas-dust disks formed from prestellar cores of various mass $M_{\rm core}$ and mass-to-magnetic flux ratio $\lambda$.  }
{Numerical magnetohydrodynamics simulations in the thin-disk limit were employed to study the long-term ($\sim 1.0$~Myr) evolution of protoplanetary disks with an adaptive turbulent $\alpha$-parameter, which depends explicitly on the strength of the magnetic field and ionization fraction in the disk. The numerical models also feature the co-evolution of gas and dust, including the back-reaction of dust on gas and dust growth.}
{Dead zone with a low ionization fraction $x \la 10^{-13}$ and temperature on the order of several hundred Kelvin forms in the inner disk soon after its formation, extending from several to several tens of astronomical units depending on the model. The dead zone features pronounced dust rings that are formed due to the concentration of grown dust particles in the local pressure maxima. 
Thermal ionization of alkaline metals in the dead zone trigger the MRI and associated accretion burst, which is characterized by a sharp rise, small-scale variability in the active phase, and fast decline once the inner MRI-active region is depleted of matter. The burst occurrence frequency is highest in the initial stages of disk formation,  and is driven by gravitational instability (GI), but declines with diminishing disk mass-loading from the infalling envelope. There is a causal link between the  initial burst activity and the strength of  GI in the disk fueled by mass infall from the envelope. We found that the MRI-driven burst phenomenon occurs for $\lambda$=2--10, but diminishes in models with $M_{\rm core}\la 0.2~M_\odot$, suggesting a lower limit on the stellar mass for which the MRI-triggered burst can occur.}
{The MRI-triggered bursts occur for a wide range of mass-to-magnetic flux ratios and initial cloud core masses. The burst occurrence frequency is highest in the initial disk formation stage and reduces as the disk evolves from a gravitationally unstable to a viscous-dominated state. The MRI-triggered bursts are intrinsically connected with the dust rings in the inner disk regions, as they can be a manifestation of one and the same phenomenon -- the formation of a dead zone.}

\authorrunning{Vorobyov et al.}
\titlerunning{MRI-triggered accretion bursts}

\maketitle

\section{Introduction}
Young (sub-)solar mass protostars can experience luminosity bursts 
known as FUor eruptions, named after the first known example of this kind -- the FU Orionis system. The peak luminosity during these bursts ranges from several tens to several hundreds of solar luminosities, raising the disk temperature and making FUors an attractive tool for studying the chemical composition of the disk and dust distribution in the inner disk regions, which are otherwise hardly accessible in quiescent systems.    Since 1937 several dozens of such objects have been discovered \citep{Audard2014}, many more candidates are being monitored \citep[e.g.,][]{Pena2017}, and it is believed that the luminosity bursts are caused by a sharp increase in mass accretion from the disk onto the star. 

Several mechanisms for the accretion bursts have been proposed in the past three decades, which involve either disk instabilities of some kind or perturbations to the disk driven by external or internal agents \citep{Bonnell1992,BL1994,Armitage2001,Lodato2004,VB2005,Pfalzner2008,Zhu2009,Forgan2010,Machida2011,Martin2012,Nayakshin2012,Bae2014,Riaz2018,Kuffmeier2018}. Recent developments even suggest the existence of accretion bursts in young high-mass protostars \citep{Caratti2016,Meyer2017}, but no firm counterparts have yet been found in the intermediate-mass regime.   

The importance of accretion bursts for the evolution of the star and its circumstellar disk cannot be underestimated. The bursts raise the gas and dust temperatures, causing sublimation of species (such as H$_2$O or CH$_3$OH) that are otherwise found in the disk midplane predominantly in the icy form \citep{Cieza2016,Lee2019}.  The desorption and adsorption of ices caused by the burst can trigger a chain of reactions that can lead to enhancements in certain complex organic molecules \citep{Taquet2016,Wiebe2019}. The bursts can affect the dust growth via evaporation of icy mantles followed by collisional shattering, but also promote dust growth through preferential recondensation of water ice \citep{Hubbard2016}. Bursts can also affect the evolution of the star itself, causing stellar bloating and dramatic excursions in the Hertzsprung-Russell diagram  if a (small) fraction of the accreted energy is absorbed by the star \citep{Elbakyan2019}.

Most of these effects depend sensitively on the magnitude, duration, and frequency of the bursts that may occur in the early evolution of a young stellar object. While the burst magnitude  can be estimated from multi-waveband observations, the burst duration and frequency is difficult to derive observationally because of the long timescales of the bursts, although attempts were made to infer them from the known statistics of the bursts \citep{Scholz2013, Hillenbrand2015,Pena2019, Fischer2019}. Alternatively, the burst characteristics can  be estimated from known numerical models \citep[e.g.][]{VB2015}, which motivates further  investigations into the nature of the burst phenomenon.

In this paper, we revisit the accretion burst model that relies on triggering the magneto-rotational instability (MRI) in the inner disk regions. This scenario was developed and elaborated in a series of works using one-, two- and three-dimensional numerical hydrodynamics simulations \citep[e.g.,][]{Armitage2001,Zhu2009,Bae2014,Zhu2020,Kadam2020}.
While 3D simulations can zoom in on subtle details of the burst \citep[e.g.,][]{Zhu2020}, simplified 2D simulations can provide a panoramic view on the burst phenomenon over many model realization and long evolutionary times. 
We use the two-dimensional thin-disk models and further elaborate this scenario by introducing (in a simplified manner) magnetic fields and dust dynamics with growth. Unlike most previous simulations, we start our computations from the gravitational collapse of pre-stellar cores with different masses and mass-to-magnetic-flux ratios, which allows us to explore the dependence of the burst characteristics on the initial conditions in pre-stellar cores. Our expertise in another accretion burst model -- disk gravitational fragmentation followed by infall of the clumps \citep{VB2010,VB2015} -- allows us to make direct comparisons between different accretion burst mechanisms.

The paper is organized as follows. In Sect.~\ref{Sec:model} we provide a detailed description of our model. In Sect.~\ref{LongEvol} we describe the global evolution of the disk with a magnetic field. In Sect.~\ref{Sec:Ideal_burst} we focus on the MRI-triggered accretion bursts. The results of our parameter-space study are provided in Sect.~\ref{ParamStudy}. Comparison with other relevant work and model caveats are considered in Sect.~\ref{comparison}. The main results are summarized in Sect.~\ref{conclude}.

\section{Model description}
\label{Sec:model}

Our numerical model is a modification of the thin-disk model for the formation and long-term evolution of gaseous and dusty disks described in detail in \citet{Vorobyov2018}. In this work,
we also take magnetic fields in the flux-freezing approximation into account and implement the MRI activation based on the adaptive $\alpha$-parameterization. Below, we briefly review the model's main constituent parts and equations.

We start our numerical simulations from the gravitational collapse of a starless magnetized  cloud core with a spatially uniform mass-to-magnetic-flux ratio,
continue into the embedded phase of star formation, during which
a star, disk, and envelope are formed, and terminate our simulations when the age of the star becomes older than about 1.0~Myr and the parental core dissipates. 
Such long integration times are made possible by the use of the thin-disk approximation, the justification of which is provided in \citet{VB2010}. In our model, the core has the form of a flattened pseudo-disk, a  spatial  configuration  that  can  be  expected  in  the  presence of rotation and large-scale magnetic fields \citep[e.g.,][]{Basu1997}. As the collapse proceeds, the inner regions of the core spin up and a centrifugally balanced circumstellar disk forms  when the inner infalling layers of the core hit the centrifugal barrier near the central sink cell. The latter
is introduced at $r_{\rm sc}=0.52$~au to avoid too small time steps imposed by the Courant condition. The protostellar disk occupies the inner part of the numerical polar grid
and its outer parts are exposed to intense mass-loading from the infalling core in the initial embedded phase of evolution.
We note that our global disk formation simulations feature one of the smallest possible sink cells, allowing us to resolve the gas and dust dynamics on sub-au scales. For example, the thin-disk simulations of \citet{Bae2014} set the inner sink at 0.2~au but use a factor of 2 smaller grid zones. The smallest sink radius in global three-dimensional simulations that adopt nested grids is about 1~au \citep[e.g.,][]{Machida2010}. 

\subsection{FEOSAD code: the gaseous component}
\label{gaseous}

The main physical processes considered in the Formation and Evolution Of a Star And its circumstellar Disk (FEOSAD) code  include viscous and shock heating, irradiation by the forming star,  background irradiation with a uniform temperature set equal to the initial temperature of the natal cloud core,
radiative cooling from the disk surface, friction between the gas and dust components, self-gravity of gaseous and dust disks, and magnetic field pressure and tension. Ohmic dissipation and ambipolar diffusion will be taken into account in a follow-up study.  The code is written in the thin-disk limit, complemented by a calculation of the gas vertical  scale height using an assumption of local hydrostatic equilibrium as described in \citet{VB2009}. The resulting  model has a flared structure (because the disk vertical scale height increases with radius), which guarantees that both the disk and envelope receive a fraction of the irradiation energy  from the central protostar. The pertinent equations of mass, momentum, and energy transport for the gas component are
\begin{equation}
\label{cont}
\frac{{\partial \Sigma_{\rm g} }}{{\partial t}}   + \nabla_p  \cdot 
\left( \Sigma_{\rm g} \bl{v}_p \right) =0,  
\end{equation}
\begin{eqnarray}
\label{mom}
\frac{\partial}{\partial t} \left( \Sigma_{\rm g} \bl{v}_p \right) &+&  [\nabla \cdot \left( \Sigma_{\rm
g} \bl{v}_p \otimes \bl{v}_p \right)]_p  =   - \nabla_p {\cal P}  + \Sigma_{\rm g} \, \left( \bl{g}_p +\bl{g}_\ast \right) + \nonumber
\\ 
 &+& (\nabla \cdot \mathbf{\Pi})_p  - \Sigma_{\rm d,gr} \bl{f}_p 
+  {B_z {\bl B}_p^+ \over 2 \pi}
- H_{\rm g}\, \nabla_p \left({B_z^2 \over 4 \pi}\right),
\end{eqnarray}
\begin{equation}
\frac{\partial e}{\partial t} +\nabla_p \cdot \left( e \bl{v}_p \right) = -{\cal P} 
(\nabla_p \cdot \bl{v}_{p}) -\Lambda +\Gamma + 
\left(\nabla \bl{v}\right)_{pp^\prime}:\Pi_{pp^\prime}, 
\label{energ}
\end{equation}
where the subscripts $p$ and $p^\prime$ refer to the planar components
$(r,\phi)$  in polar coordinates, $\Sigma_{\rm g}$ is the gas 
surface density,  $e$ is the internal energy per surface area,  ${\cal P}$
is the vertically integrated gas pressure calculated via the ideal  equation of state as ${\cal P}=(\gamma-1) e$ with $\gamma=7/5$, $\bl{v}_{p}=v_r
\hat{\bl r}+ v_\phi \hat{\bl \phi}$  is the gas velocity in the disk plane, $\nabla_p=\hat{\bl r} \partial / \partial r + \hat{\bl
\phi} r^{-1} \partial / \partial \phi $ is the gradient along the planar coordinates of the disk,  { and $H_{\rm g}$ is the vertical scale height of the gas disk calculated using an assumption of the local hydrostatic balance in the gravitational field of the disk and the star \citep[see][]{VB2009}. The term $\bl{f}_p$ is the drag force per unit mass between dust and gas, and $\Sigma_{\rm d,gr}$ is the gas surface density of grown dust  explained in more detail in Sect.~\ref{dustycomp}. A similar term enters the equation of dust dynamics and it is computed using the method laid out in \citet{Stoyan2018}. Finally, $B_z$ is the vertically  constant  but radially and azimuthally varying $z$-component of the magnetic field  within the disk thickness
and  ${\bl B}_p^+=B_r^+ \hat{\bl r}+ B_\phi^+ \hat{\bl \phi}$ are the planar components of the magnetic field at the top surface of the disk. {We assumed the midplane symmetry, so that ${\bl B}_p^-=- {\bl B}_p^+$.   The last two terms on the
right-hand side of Equation~(\ref{mom}) are the Lorentz force, including the magnetic tension term proportional to $B_z B_p^+$ and the vertically integrated magnetic pressure gradient. The magnetic tension term arises formally from an application of the Maxwell stress tensor, but can be understood intuitively as the interaction of an electric current at the disk surface (caused by a discontinuity of $B_p$ from the interior to the outside surface) with the component $B_z$ inside the disk. 
}

The gravitational acceleration in the disk
plane,  $\bl{g}_{p}=g_r \hat{\bl r} +g_\phi \hat{\bl \phi}$, takes into account self-gravity of the gaseous and dust disk components found by solving for the disk gravitational potential using the Poisson integral 
\begin{eqnarray}
\label{potential}
   \Phi(r,\phi,z&=&0)  =  \\ \nonumber
   &-& G \int_{r_{\rm sc}}^{r_{\rm out}} r^\prime dr^\prime
                   \int_0^{2\pi}
                \frac{\Sigma_{\rm tot}(r^\prime,\phi^\prime) d\phi^\prime}
                     {\sqrt{{r^\prime}^2 + r^2 - 2 r r^\prime
                        \cos(\phi^\prime - \phi) }}  \, ,
\end{eqnarray}
where $r_{\rm out}$ is the size of the cloud core and $\Sigma_{\rm tot}$ is the sum of the gas and dust surface densities
\citep[see][]{BT87}. This then yields the gravitational field $\bl{g}_p = - \nabla_p \Phi(r,\phi,z=0)$ in the plane of the disk. We have assumed that all the matter density $\Sigma_{\rm tot}$ that contributes to the gravitational potential is contained within the disk. Formally, Equation (\ref{potential}) is the solution at $z=0$ to Laplace's equation $\nabla^2 \Phi(r,\phi,z)=0$ for $z \geq 0$ with boundary condition $g_z^+(r,\phi,z=0) \equiv - \lim_{z\to 0^+}\partial \Phi/\partial z = - 2 \pi G \Sigma_{\rm tot}(r,\phi)$. Once a central protostar is formed, we add its gravitational field
\begin{equation}
    g_{r,*} = - {G M_* \over r^2}
\end{equation}
to the gravitational acceleration in the disk plane $\bl{g}_{p}$,  where $M_*$ is the mass accumulated in the protostar.

The cooling and heating rates $\Lambda$ and $\Gamma$ take the disk
cooling and heating due to stellar and background irradiation
into account based on the analytical solution of the radiation transfer
equations in the vertical direction
\citep[for details see][]{Dong2016}\footnote{The cooling and heating rates in \citet{Dong2016}
are written for one side of the disk and need to be multiplied by a factor of~2.},
\begin{equation}
\Lambda=\frac{8\tau_{\rm P} \sigma T_{\rm mp}^4 }{1+2\tau_{\rm P} + 
{3 \over 2}\tau_{\rm R}\tau_{\rm P}},
\end{equation}
where $T_{\rm mp}={\cal P} \mu / {\cal R} \Sigma_{\rm g}$ is the midplane
temperature,  $\mu=2.33$ is the mean molecular weight,  $\cal R$ is the
universal  gas constant, $\sigma$ is the Stefan-Boltzmann constant, 
 $\tau_{\rm R}=0.5 \Sigma_{\rm d,tot} \kappa_R$  and $\tau_{\rm P}=0.5 \Sigma_{\rm d,tot} \kappa_{\rm P}$ are the  Rosseland and Planck
optical depths to the disk midplane, and  $\kappa_{\rm P}$ 
and $\kappa_{\rm R}$ (in cm$^{2}$~g$^{-1}$) are the Planck and Rosseland mean opacities taken from
\citet{Semenov2003}.  Here, $\Sigma_{\rm d,tot}$ is the total surface density of dust (including small and grown components). We note the the adopted opacities do not take dust growth into account.

The heating function per surface area of the disk is expressed as
\begin{equation}
\Gamma=\frac{8\tau_{\rm P} \sigma T_{\rm irr}^4 }{1+2\tau_{\rm P} + {3 \over 2}\tau_{\rm R}\tau_{\rm
P}},
\end{equation}
where $T_{\rm irr}$ is the irradiation temperature at the disk surface 
determined from the stellar and background black-body irradiation as
\begin{equation}
T_{\rm irr}^4=T_{\rm bg}^4+\frac{F_{\rm irr}(r)}{\sigma},
\label{fluxCS}
\end{equation}
where $F_{\rm
irr}(r)$ is the radiation flux (energy per unit time per unit surface
area)  absorbed by the disk surface at radial distance  $r$ from the
central star. The latter quantity is calculated as 
\begin{equation}
F_{\rm irr}(r)= \frac{L_\ast}{4\pi r^2} \cos{\gamma_{\rm irr}},
\label{fluxF}
\end{equation}
where $\gamma_{\rm{irr}}$ is the incidence angle of radiation arriving at
the disk surface (with respect to the normal) at radial distance $r$ calculated as \citep{VB2010}
\begin{equation}
\cos\gamma_{\rm irr}=\cos\alpha_{\rm irr} \cos\beta_{\rm irr} \left( \tan\alpha_{\rm irr}
- \tan\beta_{\rm irr} \right),
\end{equation}
where $\cos\alpha_{\rm irr}=dr/(dr^2+dH_{\rm g}^2)^{1/2}$, $\cos\beta_{\rm irr}=r/(r^2+H_{\rm g}^2)^{1/2}$,
$\tan\alpha_{\rm irr}=dH_{\rm g}/dr$, and $\tan\beta_{\rm irr}=H_{\rm g}/r$.} The stellar luminosity $L_{\ast}$ is the sum of the accretion luminosity  $L_{\rm {\ast,accr}}=0.5 G M_{\ast} \dot{M}
/R_{\ast}$ arising from the gravitational energy of accreted gas and the
photospheric luminosity $L_{\rm{\ast,ph}}$ due to gravitational
compression and deuterium burning in the stellar interior. The stellar
mass $M_{\ast}$ and accretion rate onto the star $\dot{M}$ are determined
using the amount of gas passing through the sink cell. 
The properties of
the forming protostar ($L_{\rm{\ast,ph}}$ and radius $R_{\ast}$) are
calculated using the stellar evolution tracks obtained with the STELLAR code of \citet{Yorke2008}. 

\subsection{Dust component}
\label{dustycomp}
In our model, dust consists of two components: small dust ($a_{\rm min}<a<a_\ast$)  and grown dust ($a_\ast \le a<a_{\rm max}$), where $a_{\rm min}=5\times 10^{-3}$~$\mu$m, $a_\ast=1.0$~$\mu$m (both are constants of time and space),  and $a_{\rm max}$ is a dynamically varying maximum radius of dust grains, which depends on the efficiency of radial dust drift and on the rate of dust growth. All dust grains are considered to be spheres with material density $\rho_{{\rm s}}=3.0\,{\rm g~cm}^{-3}$. Small dust constitutes the initial reservoir
for dust mass and is gradually converted in grown dust as the disk forms and evolves.   Small dust is assumed to be dynamically coupled to gas (because it is composed of sub-micron dust grains by definition), meaning that we only solve the continuity equation for small dust grains, while the dynamics of grown dust is controlled by friction with the gas component and by the total gravitational potential of the star, and the gaseous and dusty components. The conversion from small to grown dust is considered by calculating the dust growth rate (see  Eq.~(\ref{GrowthRate}) below).
The resulting continuity and momentum equations for small and grown dust are
\begin{equation}
\label{contDsmall}
\frac{{\partial \Sigma_{\rm d,sm} }}{{\partial t}}  + \nabla_p  \cdot 
\left( \Sigma_{\rm d,sm} \bl{v}_p \right) = - S(a_{\rm max}),  
\end{equation}
\begin{equation}
\label{contDlarge}
\frac{{\partial \Sigma_{\rm d,gr} }}{{\partial t}}  + \nabla_p  \cdot 
\left( \Sigma_{\rm d,gr} \bl{u}_p \right) = S(a_{\rm max}),  
\end{equation}
\begin{eqnarray}
\label{momDlarge}
\frac{\partial}{\partial t} \left( \Sigma_{\rm d,gr} \bl{u}_p \right) +  [\nabla \cdot \left( \Sigma_{\rm
d,gr} \bl{u}_p \otimes \bl{u}_p \right)]_p  &=&   \Sigma_{\rm d,gr} \, \left( \bl{g}_p + \bl{g}_\ast \right) + \nonumber \\
 + \Sigma_{\rm d,gr} \bl{f}_p + S(a_{\rm max}) \bl{v}_p,
\end{eqnarray}
where $\Sigma_{\rm d,sm}$ and $\Sigma_{\rm d,gr}$ are the surface
densities of small and grown dust, $\bl{u}_p$ describes the planar components of the grown dust velocity, and $S(a_{\rm max})$ is the rate 
of conversion from small to grown dust per unit surface area. In this study, we assume that dust and gas are vertically mixed, which is justified in young gravitationally unstable and/or MRI active disks \citep{Rice2004,Yang2018}.
However, in more evolved disks dust settling becomes significant. Its effect on the global evolution of gaseous and dusty disks was considered in our previous study \citep{Vorobyov2018}, showing that dust settling accelerates the conversion of small to grown dust.

The rate of small-to-grown dust conversion $S(a_{\rm max})$ is derived based on the assumption that each of the two dust populations (small and grown) has the size distribution $N(a)$ described by a simple power-law function $N(a)= C a^{-p}$ with a fixed exponent $p=3.5$ and a normalisation constant $C$. After noting than that total dust mass stays constant during dust growth, the change in the surface density of small dust due to conversion into grown dust $\Delta \Sigma_{\rm d,sm}=\Sigma_{\rm d,sm}^{n+1}-\Sigma_{\rm d,sm}^{n}$ can be expressed as
\begin{equation}
    \Delta\Sigma_{\mathrm{d,sm}} =  \Sigma_{\mathrm{tot}}^n 
    \frac
    { I_1
    \left( 
    C_{\rm sm}^{n+1} C_{\rm gr}^n \, I_2 - 
    C_{\rm sm}^n C_{\rm gr}^{n+1} I_3
    \right)
    }
    {
    \left( 
    C_{\rm sm}^{n+1} I_1+  C_{\rm gr}^{n+1} I_3
    \right)  
    \left( 
    C_{\rm sm}^{n} I_1 + C_{\rm gr}^{n} I_2
    \right)
    \label{growth1}
    },
\end{equation}
where $C_{\rm sm}$ and $C_{\rm gr}$ are the normalisation constants for small and grown dust size distributions at the current ($n$) and next ($n+1$) time steps,   and the integrals $I_1$, $I_2$, and $I_3$ are defined as
\begin{equation}
   I_1= \int_{a_{\rm min}}^{a_*} a^{3-\mathrm{p}}da, \,\,\,\,
    I_2=\int_{a_*}^{a_{\mathrm{max}}^{n}} a^{3-\mathrm{p}}da,
    \,\,\,\,
    I_3=\int_{a_*}^{a_{\mathrm{max}}^{n+1}} a^{3-\mathrm{p}}da.
\end{equation}

By introducing different normalisation constants for small and grown dust we implicitly assume that the dust size distribution can be discontinuous at $a_\ast$, which may develop due to different dynamics of small and grown dust. In this work, we suggest that dust growth due to the $S$ term smooths out the discontinuity each time it could occur after the advection step. Physically this assumption corresponds to the dominant role of dust growth rather than dust flow in setting the shape of the dust size distribution.  This can be achieved by setting $C_{\rm sm}^{n+1}=C_{\rm gr}^{n+1}$ in Equation~(\ref{growth1}). After determining the normalization constants $C_{\rm sm}^n$ and $C_{\rm gr}^n$ from the known dust surface densities $\Sigma_{\rm d,sm}^n$ and $\Sigma_{\rm d,gr}^n$ at the current time step, the rate of small-to-grown dust conversion during one time step $\Delta t$ is then written as
\begin{equation}
\label{GrowthRate}
   S(a_{\rm max}) = - {\Delta\Sigma_{\mathrm{d,sm}} \over \Delta t} = 
    \frac
    {
    \Sigma_{\rm d,sm}^n I_3  - 
    \Sigma_{\rm d,gr}^n I_1
    }
    {
    I_4 \Delta t 
    },
\end{equation}
where the integral $I_4$ is written as
\begin{equation}
    I_4= \int_{a_{\mathrm{min}}}^{a_{\rm max}^{n+1}} a^{3-\mathrm{p}}da,
\end{equation}


To complete the calculation of $S(a_{\rm max})$, the maximum radius of grown dust $a_{\rm max}$ in a given computational cell must be computed at each time step. 
The evolution of $a_{\rm max}$ is described as
\begin{equation}
{\partial a_{\rm max} \over \partial t} + ({\bl u}_{\rm p} \cdot \nabla_p ) a_{\rm max} = \cal{D},
\label{dustA}
\end{equation}
where the growth rate $\cal{D}$ accounts for the change in $a_{\rm max}$ due to coagulation and the second term on the left-hand side account for the change of $a_{\rm max}$ due to dust flow through the cell (the left-hand side is the full derivative  of $a_{\rm max}$ over time).  We write the source term $\cal{D}$ as
\begin{equation}
\cal{D}=\frac{\rho_{\rm d} {\it v}_{\rm rel}}{\rho_{\rm s}},
\label{GrowthRateD}
\end{equation}
where $\rho_{\rm d}$ is the total dust volume density and $v_{\rm rel}$ is the
dust-to-dust collision  velocity. 
The adopted approach is similar to 
the monodisperse model of \citet{Stepinski1997} and is described in more detail in \citet{Vorobyov2018}.
The maximum value of $a_{\rm max}$ is limited  by the fragmentation barrier \citep{Birnstiel2012} defined as
\begin{equation}
 a_{\rm frag}=\frac{2\Sigma_{\rm g}v_{\rm frag}^2}{3\pi\rho_{\rm s}\alpha c_{\rm s}^2},
 \label{afrag}
\end{equation}
where  $v_{\rm frag}$ is the fragmentation
velocity set equal to a constant value of 3~m~s$^{-1}$ and $c_{\rm s}$ is the speed of sound. Whenever $a_{\rm max}$ exceeds $a_{\rm frag}$, the growth rate $\cal{D}$ is set to zero.
We note that FEOSAD does not consider the Stokes regime of dust dynamics (which requires a different approach to calculating the friction force, see \citet{Stoyan2020}) and the growth of dust is also limited to keep the size of dust particles within the Epstein regime. We also note that we do not take the possible dust sublimation at high temperatures into account and neglect dust diffusion associated with turbulence, which may affect the shape of dust rings disccused in Sect.~\ref{longdisk}. To summarise, our dust growth model allows us to determine the maximum size of dust grains and the dust-to gas mass ratio as a function of radial and azimuthal position in the disk, assuming that the slope $p$ of the dust size distribution stays locally and globally constant.

\subsection{Magnetic field calculations}
\label{magfield}
In Equation~(\ref{mom}), the last two terms represent the magnetic tension force and the force due to the magnetic pressure gradient. We neglected a smaller term  $({\bl B}_p^+)^2 \nabla_p H_{\rm g}/(4 \pi)$ that measures the planar force due to the pressure of the magnetic field at the surface pushing the disk when there is a planar gradient of its scale height, because a numerical instability of uncertain origin often develops when this term is taken into account. 

\subsubsection{The $z$-component of magnetic field}
To calculate the vertical component of magnetic field $B_z$, we consider a simplified form of the induction
equation taking Ohmic dissipation and magnetic ambipolar diffusion into account. The induction
equation can be written in a general form as (see, for example, \cite{Dudorov95})
\begin{equation}
  \frac{\partial\bl{B}}{\partial t} = \nabla\times\left(\left(\bl{v} + \bl{v}_{\rm{ad}}\right)\times \bl{B}\right)  -\nabla\times\left(\nu_{\rm{m}} \left(\nabla\times\bl{B}\right)\right), \label{Eq:Induction}
\end{equation}
where $\bl{B}$ is the magnetic field, $\bl{v}$ is the gas velocity, $\bl{v}_{\rm{ad}}$ is the velocity of ambipolar diffusion, and $\nu_{\rm{m}}$ is the Ohmic diffusivity. 
Assuming steady magnetic ambipolar diffusion, we can write ${\bl v}_{\rm ad}$ as
\begin{equation}
        {\bl v}_{\rm{ad}} = \frac{\left(\nabla\times{\bl B}\right) \times {\bl B}}{4\pi R_{\rm{in}}},\label{Eq:MADvelocity}
\end{equation}
where $R_{\rm{in}} = \mu_{\rm{in}}n_{\rm{i}}n_{\rm{n}}\langle\sigma v\rangle_{\rm{in}}$ is the friction
coefficient  between ions and neutrals, $\mu_{\rm{in}}$ is the reduced mass of ions and neutrals, $n_{\rm{i}}$ is the number density of ions, $n_{\rm{n}}$ is the number density of neutrals, $\langle\sigma v\rangle_{\rm{in}}=2\times 10^{-9}\,\mbox{cm}^3\,\mbox{s}^{-1}$ is the `slowing-down' coefficient~\citep{spitzer_book}. Ohmic diffusivity is calculated as
\begin{equation}
        \nu_{\rm{m}} = \frac{c^2}{4\pi\sigma_{\rm{e}}},\label{Eq:eta_od}
\end{equation}
where 
\begin{equation}
        \sigma_{\rm{e}} = \frac{e^2n_{\rm{e}}}{m_{\rm{e}}\nu_{\rm{en}}},\label{Eq:sigma_e}
\end{equation}
is the electrical conductivity, $e$ is the electron charge, $n_{\rm{e}}$ is the number density of electrons, $m_{\rm{e}}$ is the mass of the electron, and $\nu_{\rm{en}}$ is the mean collision rate between electrons and neutral particles, defined as
\begin{equation}
        \nu_{\rm{en}} = \langle\sigma v\rangle_{\rm{en}} n_{\rm{n}},\label{Eq:nu_en}
\end{equation}
where $\langle\sigma v\rangle_{\rm{en}}= 10^{-7}\,\rm{cm}^3\,\rm{s}^{-1}$ \citep{Nakano1984} is the `slowing-down' coefficient for the electron-neutral collisions.


In the adopted thin-disk approximation, the magnetic field and gas velocity have the following non-zero components in the disk: $\bl{B}=\left(0,\,0,\, B_z\right)$ and $\bl{v}=\left(v_r,\, v_{\varphi},\, 0\right)$, respectively. The $z$-component of Equation~(\ref{Eq:Induction}) can then be written as
\begin{eqnarray}
\frac{\partial B_z}{\partial t} &=& -\frac{1}{r}\frac{\partial}{\partial r}\left(rv_rB_z\right) - \frac{1}{r}\frac{\partial}{\partial \varphi}\left(v_{\varphi}B_z\right) + \nonumber \\
 & & \frac{1}{r}\frac{\partial}{\partial r}\left(r\eta\frac{\partial B_z}{\partial r}\right) + \frac{1}{r}\frac{\partial}{\partial \varphi}\left(\eta\frac{1}{r}\frac{\partial B_z}{\partial \varphi}\right), \label{Eq:Bz}
\end{eqnarray}
where 
\begin{equation}
        \eta = \nu_{{\rm m}} + \eta_{\rm{mad}}\label{Eq:eta_tot}
\end{equation}
is the total magnetic diffusivity, and
\begin{equation}
\eta_{\rm{mad}} = \frac{B_z^2}{4\pi R_{\rm{in}}}\label{Eq:eta_mad}
\end{equation}
is the ambipolar diffusivity. The first two terms in the right-hand side of Equation~(\ref{Eq:Bz}) are responsible for the advection of the $z$-component of  magnetic field, while the last two terms describe the magnetic field diffusion.

To solve numerically Equation~(\ref{Eq:Bz}), we use operator splitting in physical processes and split
it into advection and  diffusion equations:
\begin{equation}
\frac{\partial B_z}{\partial t} = -\frac{1}{r}\frac{\partial}{\partial r}\left(rv_rB_z\right) - \frac{1}{r}\frac{\partial}{\partial \varphi}\left(v_{\varphi}B_z\right),
\label{Eq:AdvectBz}
\end{equation}
\begin{equation}
\frac{\partial B_z}{\partial t} = \frac{1}{r}\frac{\partial}{\partial r}\left(r\eta\frac{\partial B_z}{\partial r}\right) + \frac{1}{r}\frac{\partial}{\partial \varphi}\left(\eta\frac{1}{r}\frac{\partial B_z}{\partial \varphi}\right). \label{Eq:Bz_d}
\end{equation}
Equation~(\ref{Eq:AdvectBz}) is solved using the same third-order accurate piecewise-parabolic 
advection scheme \citep{Colella1984} as for other hydrodynamic quantities (e.g., surface density, internal energy). Equation~(\ref{Eq:Bz_d}) is a non-linear diffusion equation and its solution usually requires time-consuming implicit methods. We tried a fully implicit scheme and found that the simulations were decelerated by almost a factor of 10.  Since in this work we aimed at studying the burst statistics over long evolutionary times ($\sim$~1.0 Myr), we decided to adopt the ideal MHD limit and neglect the effect of Ohmic dissipation and ambipolar diffusion on $B_z$.
We will return to the non-ideal effects in a follow-up study focusing more on the characteristics of individual bursts.

\subsubsection{The planar components of magnetic field}
We calculate the planar components of magnetic field, $\bl{B}_p^+ = B_r^+ \hat{\bl r} +B_\phi^+ \hat{\bl \phi}$,
by solving the same Poisson integral as for the gravitational field but with the source term $-G \Sigma_{\rm tot}$ replaced with
$(B_z - B_0)/(2 \pi)$. Here, $B_0$ is the strength of a uniform background vertical magnetic field. This approach can be justified as follows.
The magnetic field in the region $z \geq 0$ can be written as 
$\bl{B} = \bl{B}' + B_0\hat{\bl z}$, where $B_0\hat{\bl z}$ represents
the background uniform magnetic field that is generated by distant currents, whereas $\bl{B}'$ is generated by currents contained within the disk. 
In the external medium to the disk, the current $\bl{j} = c/(4\pi)\nabla \times \bl{B} = 0$, hence we can write the magnetic field as the gradient of a scalar magnetic potential $\Phi_{\rm M}$.
Furthermore the divergence-free condition on $\bl{B}$ means that $\Phi_{\rm M}(r,\phi,z)$ can be determined by solving the Laplace equation $\nabla^2\Phi_{\rm M}=0$ for $z \geq 0$. For simplicity, we use the magnetic potential method to calculate $\bl{B}' = - \nabla \Phi_{\rm M}$ since $B_0\hat{\bl z}$ independently satisfies the curl-free and divergence-free conditions, and because $\bl{B}' \to 0$ as $r,z \to \infty$, in analogy with the gravitational field problem. The boundary condition at the disk upper surface is $B_z^{'+} (r,\phi,z=0) \equiv - \lim_{z\to 0^+} \partial \Phi_{\rm M}/\partial z$. Hence, the boundary value problem to determine the magnetic potential $\Phi_{\rm M}$ is formally identical to that for the gravitational potential $\Phi$ but with the source term $-2 \pi G \Sigma_{\rm tot}(r,\phi)$ replaced by $B_z^{'+}(r,\phi) = B_z^+(r,\phi) - B_0$. Consequently
\begin{eqnarray}
   \Phi_{\rm M}(r,\phi,&z&=0) = \\ \nonumber  
   &+&   \frac{1}{2\pi} \int_{r_{\rm sc}}^{r_{\rm out}} r^\prime dr^\prime
                   \int_0^{2\pi}
                \frac{[B_z^+(r^\prime,\phi^\prime)-B_0] d\phi^\prime}
                     {\sqrt{{r^\prime}^2 + r^2 - 2 r r^\prime
                        \cos(\phi^\prime - \phi) }}  \, ,
\label{potential_mag}
\end{eqnarray}
and we can then determine $\bl{B}_p^+ = \bl{B}_p^{'+} = - \nabla_p \Phi_{\rm M}(r,\phi,z=0)$ at the top surface of the sheet.

The gas that flows into the sink cell carries a magnetic flux with it. To estimate the magnitude of the $r$-component of magnetic field that is brought to the sink cell with this flow, $B_{r,{\rm sink}}$, we employ a similar method as above.  We first calculate the gravitational potential in the plane of the disk created by the matter that is accumulated in the sink cell:
\begin{eqnarray}
\Phi_{\rm sink}(r,\phi,z=0) &=& - {4 G M_{\rm sink} \over \pi r_{\rm sc}} \times \\ \nonumber
&& \left[ K_2 {r \over r_{\rm sc}} +
K_1 \left( {r_{\rm sc} \over r} - {r \over r_{\rm sc}} \right)
\right],
\end{eqnarray}
where $M_{\rm sink}$ is the total mass of gas and dust in the sink, and $K_1$ and $K_2$ are the elliptic integrals of the first and second kind. The scalar magnetic potential of the sink cell is then calculated as
\begin{equation}
    \Phi_{\rm M,sink}(r,\phi,z=0)= -\Phi_{\rm sink}   \left( {B_{z,{\rm sink}} - B_{ 0}  \over 2\pi G \Sigma_{\rm sink}} \right), 
\end{equation}
where $B_{z,{\rm sink}}$ is the magnetic field in the sink cell  and $\Sigma_{\rm sink}$ is the surface density of gas and dust in the sink cell determined from the boundary conditions (see Sect.~\ref{Boundary}). The planar component of magnetic field in the disk $\bl{B}_{\rm p}^+$ is then modified to take an additional input from the sink cell into account:  $\bl{B}_p^+ = - \nabla_p \Phi_{\rm M}(r,\phi,z=0) - \nabla_p \Phi_{\rm M,sink}(r,\phi,z=0)$.

\subsubsection{Ionization fraction}
\label{Sec:Ion}
Ohmic diffusivity (Equation \ref{Eq:eta_od}) and ambipolar diffusivity (Equation \ref{Eq:eta_mad}) depend on the ionization fraction $x=n_{\rm{e}}/(n_{\rm{e}}+n_{\rm{i}}+n_{\rm{n}})$, where we assume that electrons and ions are the only charge carriers. We calculate $x$ following~\citet{Dudorov87} (see also~\cite{Dudorov2014}). Under conditions of low temperatures, $T\lesssim 1000$~K, the ionization fraction is determined from the balance equation of collisional ionization, radiative recombination,  and recombination on dust grains
\begin{equation}
(1-x)\xi = \alpha_{\rm{r}} x^2n_{\rm{n}} + \alpha_{\rm{d}} xn_{\rm{n}},\label{Eq:xs}
\end{equation}
where $\xi$ is the ionization rate, 
\begin{equation}
\alpha_{\rm{r}} = 2.07\times 10^{-11}T^{-\frac{1}{2}}\,\rm{cm}^3\,\rm{s}^{-1}
\end{equation}
is the radiative recombination rate~\citep{spitzer_book}.

We consider ionization by cosmic rays with a rate
\begin{equation}
\xi_{\rm{cr}} = 1.0\times 10^{-17}\exp{\left(-\Sigma/\Sigma_{\rm{cr}}\right)}\,\rm{s}^{-1},
\end{equation}
where $\Sigma_{\rm{cr}}=100\,\rm{g}\,{cm}^{-2}$ is the attenuation depth of the cosmic rays. Additionally, the ionization by radionuclides is included, $\xi_{\rm{re}}=7.6\times 10^{-19}\,\rm{s}^{-1}$~\citep{umebayashi09}. The total ionization rate is $\xi = \xi_{\rm{cr}} + \xi_{\rm{re}}$.

In the regions where the gas temperature exceeds several hundred Kelvin, thermal ionization of alkali metals takes place. We evaluate the degree of thermal ionization by considering ionization of potassium as a metal with lowest ionization potential~\citep[see][]{Dudorov2014},
\begin{eqnarray}
    x_{{\rm th}} &=& 1.8\times 10^{-11}\left(\frac{T_{\rm mp}}{1000\,{\rm K}}\right)^{3/4}\left(\frac{X_{{\rm K}}}{10^{-7}}\right)^{1/2} \nonumber \\
    & & \times \left(\frac{n_{{\rm n}}}{10^{13}\,\rm{cm}^{-3}}\right)^{1/2}\frac{\exp{\left(-25000/T_{\rm mp}\right)}}{1.15\times 10^{-11}},
    \label{x_thermal}
\end{eqnarray}
where $X_{{\rm K}}$ is the cosmic abundance of potassium,  set equal to $10^{-7}$ in this work. This value is added to the ionization fraction $x$ determined from Equation~(\ref{Eq:xs}).

Our model has two populations of dust grains: small and grown ones. The total rate of recombination onto the dust grains is calculated as a sum of the rates for both populations, $\alpha_{\rm{d}} = \langle \alpha_{{\rm d}} \rangle^{{\rm small}} + \langle \alpha_{{\rm d}} \rangle^{{\rm grown}}$. The rate of recombinations onto the dust grains of a given population is calculated as
\begin{equation}
\langle \alpha_{{\rm d}} \rangle = \langle X_{{\rm d}} \cdot \sigma_{{\rm d}} \cdot v_{{\rm i}} \rangle,
\end{equation}
where $X_{{\rm d}} = n_{{\rm d}} / n_{{\rm g}}$ is the ratio of the dust volume number density $n_{{\rm d}}$ to the gas volume number density $n_{{\rm g}}$ in the disk midplane, $\sigma_{{\rm d}} = \pi a^2$ is the grain cross-section, $v_{{\rm i}}$ is the thermal speed of ions with mass $30\,m_{{\rm H}}$, which is an approximate value that is representative of the dominant ionic species, e.g., HCO$^{+}$, N$_2$H$^+$. 
We note that $X_{\rm d}$ is non-homogeneous in our model and may change spatially and temporally as the dust grows and drifts inwards.
Brackets $\langle...\rangle$ denote averaging over the dust size distribution of a given dust population (small or grown). 
The gas and dust surface densities are turned into volume
number densities in the disk midplane assuming a Gaussian distribution in the vertical direction.
Small grains are considered to be well-mixed with gas and have the same scale height as the gas. The grown dust grains are prone to sedimentation, so that their scale height can differ from the scale height of the gas. In the present simulations, we neglect this difference and assume similar scale heights for grown dust and gas. The effect of different scale heights of gas and dust was explored in \citet{Vorobyov2018}, who showed that the presence of dust settling
has little effect on the size of dust grains in the inner disk
where the grain size is limited by the fragmentation barrier,
which does not depend on the dust volume density.

\subsection{Adaptive turbulent viscosity and `dead' zones}
\label{sec:adaptive_alpha}
Viscosity is an important mechanism of mass and angular momentum transport. A possible source of viscosity in protoplanetary disks is turbulence induced by the magnetorotational instability. In FEOSAD the turbulent viscosity is taken into account via the viscous stress tensor  $\mathbf{\Pi}$, the expression for which in the thin-disk limit can be found in \citet{VB2010}. We describe the
magnitude of kinematic viscosity $\nu=\alpha c_{\rm s} H_{\rm g}$  using the $\alpha$-parameterization of \citet{SS1973}.

Unlike our previous studies of protoplanetary disks that assumed a spatially uniform and temporally constant $\alpha$-parameter \citep[e.g.,][]{Vorobyov2018}, we employed in this paper an adaptive $\alpha$-prescription based on the method presented in \citet{Bae2014}. These authors introduced an adaptive $\alpha$-parameter of the form
\begin{equation}
\label{alphavalue}
    \alpha = \frac{\Sigma_{{\rm MRI}} \, \alpha_{{\rm MRI}} + \Sigma_{{\rm dz}} \, \alpha_{{\rm dz}}}{\Sigma_{{\rm MRI}} + \Sigma_{{\rm dz}}},
\end{equation}
where $\Sigma_{{\rm MRI}}$ is the gas column density of the MRI-active layer at a given position in the disk, $\alpha_{{\rm MRI}}$ is the $\alpha$-parameter in the active layer, $\Sigma_{{\rm dz}}$ is the gas column density of the MRI-inactive layer at the same position, and $\alpha_{{\rm dz}}$ is the $\alpha$-parameter in the MRI-inactive layer. In this fashion, the adaptive $\alpha$-parameter is calculated as a density-weighted average between an MRI-active value $\alpha_{{\rm MRI}}$, usually taken to be 0.01, and an MRI-dead value $\alpha_{{\rm dz}}$, usually taken to be in the $10^{-4}$--$10^{-5}$ range. The spatial region characterized by a significantly reduced $\alpha$-value is called the dead zone. In the method of \citet{Bae2014}, the gas column density
of the MRI-active layer of the disk is a fixed parameter, usually taken to be 100~g~cm$^{-2}$. If the gas surface density to the midplane of the disk is smaller than this value, the entire disk column is MRI-active. In this method, the transition from the MRI-dead to MRI-active state can also occur if the gas temperature in the midplane exceeds a certain critical value for thermal ionization to set in, also usually taken to be a parameter in the 1000--1500~K range.

In this paper, we further upgraded this method with the purpose to reduce the number of free parameters that enter the adaptive $\alpha$-prescription. More specifically, the
dead zone is now defined as a disk region characterized by small ionization fraction and effective diffusion of magnetic field. In these disk regions, the MRI is not supposed to develop and, as a result, the MHD turbulence is weakened or absent. The boundary of the dead zone approximately coincides with the isosurface of the ionization fraction, at which the condition for suppressing the MRI is fulfilled (see Gammie, 1996). Following~\citet{Sano2000}, we determine the boundary of the dead zone from the equality of the wavelength of the most unstable MRI mode, $\lambda_{{\rm MRI}}$, and the gas scale height of the disk $H_{\rm g}$.  In the case of efficient magnetic diffusion, $\lambda_{{\rm MRI}} = 2\pi \eta /v_{{\rm a}}$, where $v_{{\rm a}}$ is the Alfv{\' e}n speed.

Following \citet{Bae2014}, we set  $\alpha_{{\rm dz}}=10^{-5}$. Unlike these authors, however, we assume that $\alpha_{{\rm MRI}}$ is not constant throughout the disk. 
The peak value of $\alpha$-parameter is higher in the disk regions directly involved in the burst than in otherwise MRI-active regions (e.g., outer disk with high ionization). This choice is motivated by numerical magnetohydrodynamics simulations of \citet{Zhu2020} suggesting that the $\alpha$-value in the innermost disk regions during the MRI burst can exceed notably the typical value of 0.01 for MRI-active disks in the non-burst state \citep{Yang2018}.

The gas column densities of the MRI-active and MRI-dead zones $\Sigma_{\rm MRI}$ and $\Sigma_{\rm dz}$ are calculated as follows. We substitute the expression for  Ohmic diffusivity (eq. \ref{Eq:eta_od}) { together with formulae (\ref{Eq:sigma_e}-\ref{Eq:nu_en}) for conductivity and collision rate with $n_{\rm e}=xn_{\rm n}$, as well as} the Alfv{\' e}n speed, $v_{{\rm a}} = B_z / \sqrt{4\pi\rho_{\rm crit}}$, with the critical gas volume density $\rho_{\rm crit}=\Sigma_{{\rm crit}} / (\sqrt{2 \pi } H_{\rm g}$) into the equality $2\pi \nu_{\rm m} /v_{{\rm a}} = H_{\rm g}$, and derive the critical gas surface density $\Sigma_{{\rm crit}}$ for the MRI development as a function of 
$H_{\rm g}$, $x$, and $B_z$,
\begin{equation}
    \Sigma_{{\rm crit}} = \left[\left(\frac{\pi}{2}\right)^{1/4}\frac{c^2m_{\rm e}\langle\sigma v\rangle_{\rm en}}{e^2}\right]^{-2}B_z^2 H_{\rm g}^3 x^2.
    \label{Eq:DZ}
\end{equation}
We note that $\Sigma_{\rm crit}$ is derived using Ohmic diffusivity but neglecting other non-ideal MHD effects. The reason for this is that Ohmic dissipation is known to prevail over ambipolar diffusion in the innermost several au of the disk \citep[see, e.g.,][]{Balbus2001,Kunz2004}, where the MRI-triggered bursts are localized. The Hall effect may also be important in these inner regions but its implementation in the thin-disk limit is not straightforward.  
The surface densities of the MRI-active and MRI-dead zones are then chosen according to the following algorithm:
\begin{eqnarray}
\label{Sigma_crit}
 \mathrm{if}\,\, \Sigma_{\rm g} &<& \Sigma_{\rm crit}   \,\, \mathrm{then} \,\,
\Sigma_{\rm MRI}=\Sigma_{\rm g}~\mathrm{and}~\Sigma_{\rm dz} = 0, \\
\mathrm{if}\,\, \Sigma_{\rm g} &\ge& \Sigma_{\rm crit} \,\, \mathrm{then} \,\, \Sigma_{\rm MRI}=\Sigma_{\rm crit} \,\, \mathrm{and} \,\, \Sigma_{\rm dz}=\Sigma_{\rm g} - \Sigma_{\rm crit}. 
\end{eqnarray}
If $\Sigma_{\rm g}\ge \Sigma_{\rm crit}$, a dead zone begins to develop at a given location in the disk. The depth of the dead zone in terms of the $\alpha$-value is determined by the balance between $\Sigma_{\rm MRI}$ and $\Sigma_{\rm dz}$ (see Eq.~\ref{alphavalue}). The analysis of Equation~(\ref{Sigma_crit}) demonstrates that a sharp increase in $\Sigma_{\rm crit}$ and hence the onset of the burst occurs if the ionization fraction $x$ experiences a sharp rise. The $B_z$-component of magnetic field and the gas scale height $H_{\rm g}$ are not expected to vary on short timescales in our model. We note that the inclusion of Ohmic dissipation (neglected in the current paper) in the magnetic induction equation may reduce the magnitude of $B_z$ in the inner disk, thus reducing $\Sigma_{\rm crit}$ and increasing the radial extent of the dead zone.
 We note that Criterion~(\ref{Eq:DZ}) formally implies that there is no MRI if $B_z$ is zero. This state is, however, never realized in our global disk simulations because some nonzero $B_z$ component is always dragged in to the disk by the cloud core collapse. Although the development of MRI is possible in a purely toroidal field geometry, the $B_z$ component is known to enhance the instability \citep[e.g.,][]{Hawley1995}.

When compared to the method of \citet{Bae2014}, $\Sigma_{\rm MRI}$ in our model is no longer constant but is a spatially and temporally varying function of $B_z$, $H_{\rm g}$, and $x$. There is also no fixed temperature for thermal activation of the MRI; the MRI thermal activation now depends on the ionization fraction $x_{\rm th}$. A similar dependence of $\Sigma_{\rm MRI}$ on the disk parameters was adopted in one-dimensional viscous disk models of \citet{Martin2012a}.

\subsection{Boundary conditions}
\label{Boundary}
The inner computational boundary cannot be placed at its physical distance from the protostar (usually much
less than 0.1 au) because of strict limitations imposed by the Courant condition on the time step. Placing the inner boundary at several au and thus relaxing the Courant condition, as was done in our previous hydrodynamic collapse simulations \citep[e.g.,][]{VB2010}, carries a risk of cutting out the disk regions that may be dynamically important for the entire disk evolution \citep[see, e.g.,][]{Vorobyov2019}.  Here, the inner boundary of the computational domain
was placed at $r_{\rm sc} = 0.52$~au, which is a reasonable compromise between computational time and physical realism, allowing us to 
capture the MRI triggering that occurs at sub-au scales. 

Care should be taken when choosing the type of flow through the inner boundary.  If
the inner boundary allows for matter to flow only in one direction, i.e., from the disk to the sink cell, then any wave-like motions near the inner boundary, such as
those triggered by spiral density waves in the disk, would result in a disproportionate flow through the sink-disk interface. As a result, an artificial depression in the gas density near the inner boundary develops in the course
of time because of the lack of compensating back flow from the sink to the disk.

To reduce the possible negative effect of the disproportionate flow, the FEOSAD code features the special inflow-outflow inner boundary condition at the sink cell-disk interface as described in \citet{Vorobyov2018} and \citet{Kadam2019}.
The mass of material that crosses the sink-disk interface is further redistributed
between the central protostar and the sink cell as $\Delta M_{\rm
flow}=\Delta M_\ast + \Delta M_{\rm sink}$ according to the following algorithm ({note that $\Delta M_{\rm flow}$ is always positive definite):
\begin{eqnarray}
 \mathrm{if}\,\, \Sigma_{\rm sink}^n < \overline{\Sigma}_{\rm in.disk}^n\,\, \mathrm{and}  \,\, v_r(r_{\rm sink})<0   \,\, \mathrm{then} \nonumber\\
 \Sigma_{\rm sink}^{n+1}=\Sigma_{\rm sink}^n&+&\Delta M_{\rm sink}/S_{\rm sink} \nonumber\\
 M_\ast^{n+1}=M_\ast^n&+&\Delta M_\ast \nonumber \\
 B_{z,{\rm sink}}^{n+1}=B_{z,{\rm sink}}^n &+& \Delta \Phi_{B,z}/S_{\rm sink} \nonumber \\
 \mathrm{if}\,\, \Sigma_{\rm sink}^n < \overline{\Sigma}_{\rm in.disk}^n\,\, \mathrm{and}  \,\, v_r(r_{\rm sink})\ge 0  \,\, \mathrm{then} \nonumber\\
 \Sigma_{\rm sink}^{n+1}=\Sigma_{\rm sink}^n&-&\Delta M_{\rm flow}/S_{\rm sink} \nonumber\\
 M_\ast^{n+1}=M_\ast^n && \nonumber \\
  B_{z,{\rm sink}}^{n+1}=B_{z,{\rm sink}}^n &-& \Delta \Phi_{B,z}/S_{\rm sink} \nonumber \\
 \mathrm{if}\,\, \Sigma_{\rm sink}^n \ge \overline{\Sigma}_{\rm in.disk}^n\,\, \mathrm{and} \,\, v_r(r_{\rm sink})<0 \,\, \mathrm{then} \nonumber\\
 \Sigma_{\rm sink}^{n+1}= \Sigma_{\rm sink}^n && \nonumber\\
 M_\ast^{n+1}= M_\ast^n &+& \Delta M_{\rm flow} \nonumber \\
  B_{z,{\rm sink}}^{n+1}=B_{z,{\rm sink}}^n && \nonumber \\
  \mathrm{if}\,\, \Sigma_{\rm sink}^n \ge \overline{\Sigma}_{\rm in.disk}^n\,\, \mathrm{and} \,\, v_r(r_{\rm sink})\ge0 \,\, \mathrm{then} \nonumber\\
 \Sigma_{\rm sink}^{n+1}= \Sigma_{\rm sink}^n &-& \Delta M_{\rm flow}/S_{\rm sink}  \nonumber\\
 M_\ast^{n+1}= M_\ast^n && \nonumber \\
  B_{z,{\rm sink}}^{n+1}=B_{z,{\rm sink}}^n &-& \Delta \Phi_{B,z}/S_{\rm sink} \nonumber 
\end{eqnarray}
Here, $\Sigma_{\rm sink}$ is the surface density of gas (or dust) in the
sink cell,  $\overline\Sigma_{\rm in.disk}$ is the averaged surface density
of gas (or dust) in the inner active disk (the averaging is usually done over
one au immediately adjacent to the sink cell), $S_{\rm sink}$ is the surface area of the sink cell, and $v_r(r_{\rm sink})$ is the radial component of velocity at the sink--disk interface. We note that $v_r(r_{\rm sink})<0$ when the gas flows from the active disk to the sink cell and $v_r(r_{\rm sink})>0$ in the opposite case.
The superscripts $n$ and $n+1$ denote the current and the updated (next time step) quantities. The exact partition between $\Delta M_\ast$ and $\Delta M_{\rm sink}$ is usually set to 95\%:5\%. This means that most of the matter that crosses the sink--disk interface quickly lands on the star, implying a fast mass transport through the sink cell which can be expected in the MRI-active limit.

The flow of matter to and from the sink cell carries magnetic flux with it. Therefore, the inner boundary condition also modifies the $z$-component of magnetic field in the sink cell $B_{z,{\rm sink}}$ based on the amount of magnetic flux $\Delta \Phi_{B,z}$ that is carried in or out of the sink cell with the flow of matter.
Our inner boundary is designed so that  an initially  spatially constant mass-to-flux ratio $\lambda$ stays constant both in the entire computational domain and in the sink cell during disk evolution in the ideal MHD limit.  The fact that $\lambda$ stays constant in the ideal MHD limit serves as an assuring test on our numerical scheme and boundary conditions. Of course, when the non-ideal MHD effects are considered, the mass-to-flux ratio may become non-homogeneous. 
{ We note that the constant $\lambda$ condition implies that most of the initial magnetic flux is advected with the gas flow to the sink cell by the end of numerical simulations. If this flux is further brought to the star, then the magnetic field of the star would become much greater than what is measured even for the most magnetized T~Tauri stars, signifying the importance of non-ideal MHD effects in redistributing the magnetic flux in the star and disk formation process.}


The calculated surface density and $z$-component of magnetic field in the sink cell $\Sigma_{\rm
sink}^{n+1}$ and $B_{z,{\rm sink}}^{n+1}$ are used at the next time step as the inner boundary values for $\Sigma_{\rm g}$, $\Sigma_{\rm d,gr}$, $\Sigma_{\rm d,sm}$, and $B_z$.
The radial velocity and internal energy at the inner boundary
are determined from the zero gradient condition, while the azimuthal velocity is extrapolated from the active disk to the sink cell assuming a Keplerian rotation.
We ensure that our
boundary conditions conserve the total mass and magnetic flux budget in the system.
Finally, we note that the outer boundary condition is set to a standard free outflow, allowing for material to flow out of the computational domain, but not allowing  any material to flow in.

\subsection{Initial conditions}
\label{Sec:IC}

The initial radial profile of the gas surface density $\Sigma_{\rm g}$ and
angular velocity $\Omega$ of the pre-stellar core has the
following form: 
\begin{equation}
\Sigma_{\rm g}=\frac{r_{0}\Sigma_{0}}{\sqrt{r^{2}+r_{0}^{2}}},\label{eq:sigma}
\end{equation}
\begin{equation}
\Omega=2\Omega_{0}\left(\frac{r_{0}}{r}\right)^{2}\left[\sqrt{1+\left(\frac{r}{r_{0}}\right)^{2}}-1\right],\label{eq:omega}
\end{equation}
where $\Sigma_{0}$ and $\Omega_{0}$ are the angular velocity and
gas surface density at the center of the core, $r_{0}=\sqrt{A}c_{\mathrm{s}}^{2}/\pi G\Sigma_{0}$
is the radius of the central plateau, where $c_{\mathrm{s}}$ is the
initial sound speed in the core. This radial profile
is typical of pre-stellar cores with a supercritical mass-to-flux ratio that are
formed through ambipolar diffusion, with the specific angular momentum
remaining constant during axially-symmetric core collapse \citep{Basu1997}. 
The value of the positive density perturbation \textit{A} is
set to 2.0, making the core unstable to collapse. 
Initially all dust is in the form of small sub-micron dust grains. The surface density and angular velocity profiles of small dust follow that of gas with the initial dust-to-gas mass ratio set equal to 1:100.

\begin{table*}
\center
\caption{\label{table1}Model parameters}
\begin{tabular}{ccccccccc}
 &  &  &  &  &  &  &  &     \tabularnewline
\hline 
\hline 
Model & $M_{\mathrm{core}}$ & $\beta$ &  $\Omega_{0}$ & $r_{0}$ & $\Sigma_{\rm g,0}$ & $r_{\mathrm{out}}$  & $\lambda$ & $M_{\rm \ast,fin}$ \tabularnewline
 & {[}$M_{\odot}${]} & {[}\%{]} &  {[}$\mathrm{km\,s^{-1}\,pc^{-1}}${]} & {[}au{]} & {[}$\mathrm{g\,cm^{-2}}${]} & {[}pc{]} & & [$M_\odot$]  \tabularnewline
\hline 
1 & 1.45 & 0.23 &  3.0 & 900 & 0.36 & 0.035 & 2.0 & 0.98 \tabularnewline
2 & 0.83 & 0.23 &  5.2 & 514 & 0.62 & 0.02 & 2.0 & 0.625  
\tabularnewline
2L & 0.83 & 0.23 &  5.2 & 514 & 0.62 & 0.02 & 10.0 & 0.608 
\tabularnewline
3 & 0.21 & 0.23 &  20.8 & 129 & 2.5 & 0.005 & 2.0 & 0.17  \tabularnewline
\hline 
\end{tabular}
\center{ \textbf{Notes.} $M_{\mathrm{core}}$ is the initial core
mass, $\beta$ is the ratio of rotational to gravitational energy,  $\Omega_{0}$ and $\Sigma_{\rm g,0}$ are the angular velocity
and gas surface density at the center of the core, $r_{0}$ is the radius
of the central plateau in the initial core,  $r_{\mathrm{out}}$ is the initial radius of the
core, $\lambda$ is the initial mass to magnetic flux ratio, and $M_{\rm \ast,fin}$ is the final stellar mass at the end of simulation.}
\end{table*}

The initial gas temperature in collapsing cores is $T_{\mathrm{init}}=20\,\mathrm{K}$. We consider several initial cloud cores, the parameters of which are listed in Table~\ref{table1}. The initial spatially uniform mass to flux ratio $\lambda=\Sigma_{\rm g}/B_z$ is given  in units of the critical value
$(2\pi \sqrt{G})^{-1}$. The initial distribution of magnetic field $B_z$ is determined from $\lambda$. The
strength of a uniform background vertical magnetic field $B_0$ is set to $10^{-5}$~Gauss.

\subsection{Solution procedure}
\label{Sec:HD}



Equations (\ref{cont})--(\ref{energ})  and (\ref{contDsmall})--(\ref{momDlarge}) are solved in polar coordinates ($r, \phi$) on a
numerical grid with $256\times256$ grid zones. The radial points are logarithmically spaced. 
The innermost grid point is located at the position of the sink cell $r_{\rm sc} = 0.52$~au, 
and the size of the first adjacent cell is about 0.018~au (it may vary slightly depending on the model). The sub-au resolution is thus achieved in the inner 30~au regions of the disk.
We use a combination of finite-differences and finite-volume methods with a time-explicit
solution procedure similar in methodology to the ZEUS code \citep{SN1992}. 
The advection is treated using the third-order-accurate piecewise-parabolic interpolation scheme of
\citet{Colella1984}.  The update of the internal energy per surface
area due to cooling $\Lambda$  and heating $\Gamma$ is done implicitly using
the Newton-Raphson method of root finding, complemented by
the bisection method where the Newton-Raphson iterations fail to converge. The accuracy is guaranteed by not allowing the internal energy to
change more than 15\% over one time step. If this condition is violated in a particular cell, we employ subcycling for this cell, i.e., the solution is sought with a local time step that is smaller
than the global time step by a factor of 2. The local time step
may be further decreased until the desired accuracy is reached.

The terms describing turbulent viscosity are implemented in the code using an explicit finite-difference
scheme. This is found to be adequate for $\alpha\le 0.01$ because other
terms (usually, the azimuthal advection) dominate in the Courant
condition that controls the time step, even though we use the FARGO algorithm to accelerate calculations in Keplerian disks \citep{Masset2000}.
A small amount of artificial viscosity is added to the code to smooth out shocks over two grid zones. The associated artificial viscosity torques
integrated over the disk area are negligible in comparison
with gravitational or turbulent viscous torques. Finally, the Courant time-step condition was modified
to include the limiters due to the magnetic field, following~\cite{SN1992b}. The adopted Courant number is 0.5. The dust friction term $f_{\rm p}$ between gas and dust (including backreaction of dust on gas) is calculated using the fully implicit algorithm presented in \citet{Stoyan2018}. The standard test problems (Sod shock tube and dusty wave) have demonstrated a clear superiority of this method over semi-implicit schemes. 


\section{Global disk evolution with magnetic field}
\label{LongEvol}
In this section we consider the global evolution of the disk in the thin-disk limit in the presence of a magnetic field. We choose model~2 as a fiducal model for this purpose.

\subsection{Initial stages of collapse and disk evolution}

\begin{figure}

\begin{centering}
\includegraphics[width=1\columnwidth]{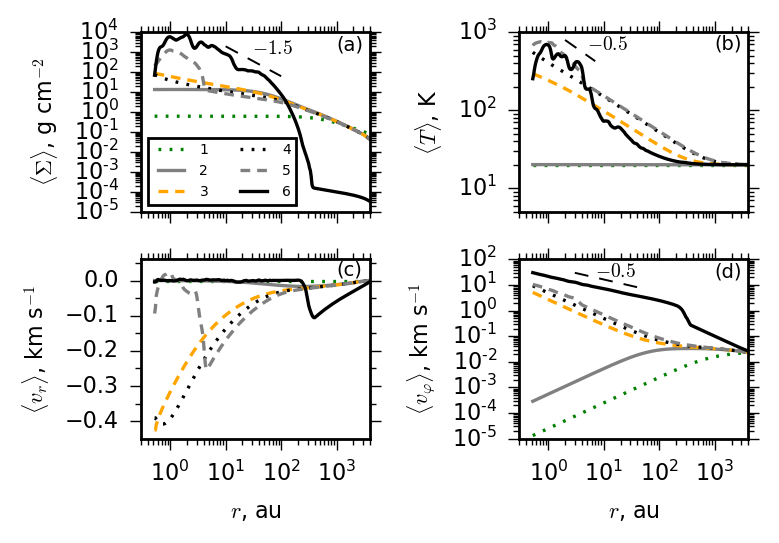}
\par\end{centering}
\caption[width=0.88\textwidth]{Azimuthally averaged radial profiles of the gas surface density (panel a), temperature (panel b), gas radial (panel c) and azimuthal velocities (panel d) at different evolution times after the onset of the simulation (1: 0~kyr, 2: 54.9~kyr, 3: 55.1~kyr, 4: 56.0~kyr, 5: 56.2~kyr, 6: 156.2~kyr). The black dashed lines with labels show typical slopes.}
\label{Fig:radial}
\end{figure}

Firstly, we discuss the process of disk formation. 
In Figure~\ref{Fig:radial}, we plot the averaged radial profiles of gas surface density $\langle \Sigma_{\rm g}\rangle$, gas midplane temperature $\langle T \rangle$, radial and azimuthal velocities of gas $\langle v_r \rangle$ and $\langle v_\phi \rangle$ at different evolution times. 
Initially, the prestellar core has a uniform, rigidly rotating plateau in the center with 
$\langle \Sigma_{\rm g} \rangle = 0.62\,\mathrm{g}\,\rm{cm}^{-2}$. In the outer parts, the surface density 
declines as $r^{-1}$, while the azimuthal velocity becomes nearly constant. The gas temperature is 20~K throughout the core. By the time $t=54.9$~kyr (grey solid lines), the surface density grows by a factor of 20 in the central part of the core. At $r>100$~au, the surface density profile obeys a nearly self-similar law $\langle \Sigma_{\rm g} \rangle \propto r^{-1}$, except for the regions near the outer boundary where a rarefaction wave develops and the surface density drops faster than $\langle \Sigma_{\rm g} \rangle
\propto r^{-1}$ \citep{VB2005b}. The gas temperature remains close to the initial value $20$~K at  this time. The radial velocity profile becomes non-monotonic: a maximum infall rate of $\langle v_r\rangle\sim 0.02\,\mathrm{km}\,\mathrm{s}^{-1}$ is reached in the outer part of the cloud, $r\approx 200$~au. The rotation profile remains nearly rigid at $r<50$~au, while differential rotation with $\langle
v_{\phi} \rangle \approx 3\times 10^{-2}$~km~s$^{-1}$ establishes in the outer regions.

\begin{figure*}[t]
\begin{center}
\includegraphics[width=0.98\textwidth, trim=0cm 0cm 0cm 0cm]{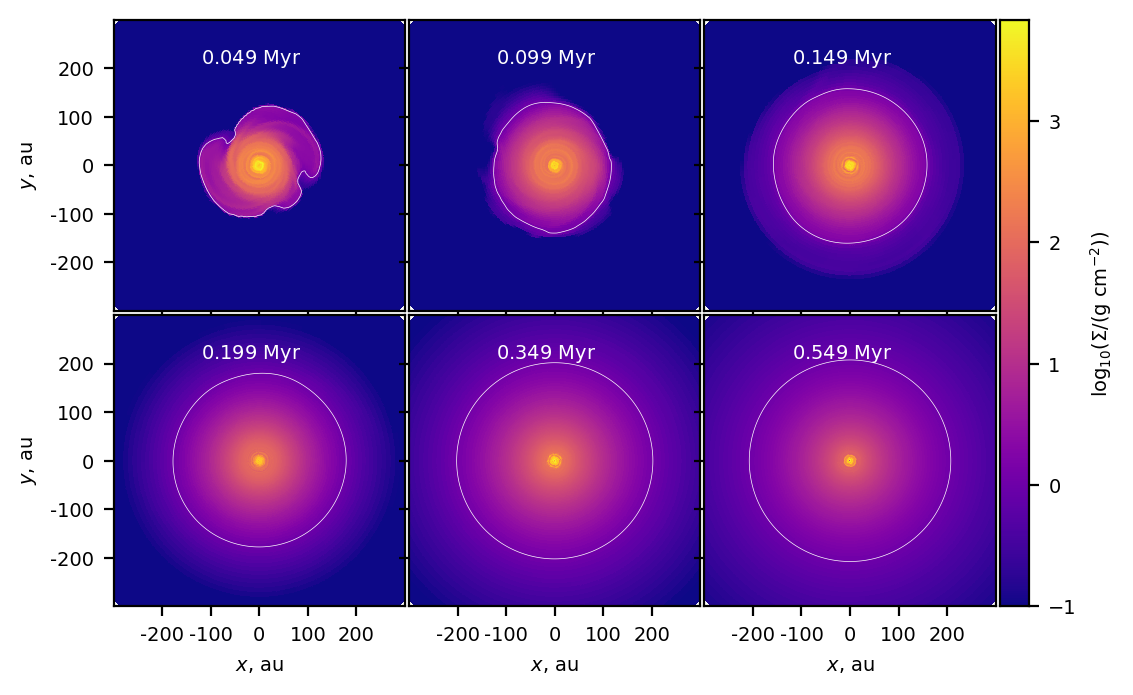}
\caption[width=0.68\textwidth]{Two-dimensional distribution of the gas surface density in inner $600\times600$~au$^2$ box at time instances $0.049$, $0.099$, $0.149$, $0.199$, $0.349$, $0.549$~Myr after disk formation. White contours mark the disk loci with $\Sigma = 1\,\mathrm{g \,cm}^{-2}$. The time is counted from the disk formation instance ($t=0.056$~Myr).}
\label{Fig:S2D}
\end{center}
\end{figure*}

At $t=55.5$~kyr after the start of the collapse (orange dashed lines), the surface density grows by two orders of magnitude and the temperature increases by factor of 10 in the central part of the core. The infall velocity $\langle v_r \rangle$ monotonically increases from nearly zero at the outer boundary of the computational domain to $-0.4\,\mathrm{km}\,\mathrm{s}^{-1}$ at the inner boundary. At this time instance, differential rotation establishes in the entire cloud with $\langle v_{\phi} \rangle$ being a decreasing function of radius.
It should be noted that the gravitational field of the system is initially determined only by the mass of the infalling gas. 

At $t\approx 55.5$~kyr, the mass of gas in the sink cell exceeded 0.02~$M_\odot$. We further assume that the phenomenon known as the second collapse, due to molecular hydrogen dissociation, has occurred in the sink cell, leading to the formation a protostar. Therefore, we place a point-like gravity source at the coordinate origin. The formation of the central star changes the gravitational field of the system, so that the gas in the inner regions starts falling faster towards the star. As a result, the gas surface density decreases at $t=56$~kyr (black dotted line) compared to the previous evolution times. The slope of the surface density profile 
is no longer self-similar. The temperature of the gas grows to $\approx 600$~K near the inner edge of the computational domain thanks to stellar illumination and compressional heating. The radial profile of $\langle v_{r} \rangle$  becomes non-monotonic: gas in the outer regions falls with increasing velocity, while the gas  in the innermost regions decelerates when 
approaching the star, ushering the formation of a centrifugally balanced disk. A maximum 
infall velocity $\approx -0.4\,\rm{km}\,\rm{s}^{-1}$ is reached at $r\approx 1$~au. 
The deceleration of the gas in the innermost region, $r<1$~au, is caused by the action of the centrifugal force. According to Figures~\ref{Fig:radial}c and \ref{Fig:radial}d, the azimuthal velocity of the gas becomes greater than the radial velocity in this region, i.e. the centrifugal barrier forms.

Shortly after the formation of the central protostar, at $t\approx 56.4$~kyr (the grey dashed lines in Figure~\ref{Fig:radial}), a bump in the gas surface density  appears at $r<5$~au. As Figure~\ref{Fig:radial}c demonstrates, the radial velocity of the gas approaches zero inside this region. This 
bump corresponds to the newly formed disk, which is the region where the centrifugal force hinders the further infall of the gas from the envelope. In the subsequent evolution, the disk grows due to the continual mass-loading from the infalling envelope.

\begin{figure}
\begin{centering}
\includegraphics[width=1\columnwidth]{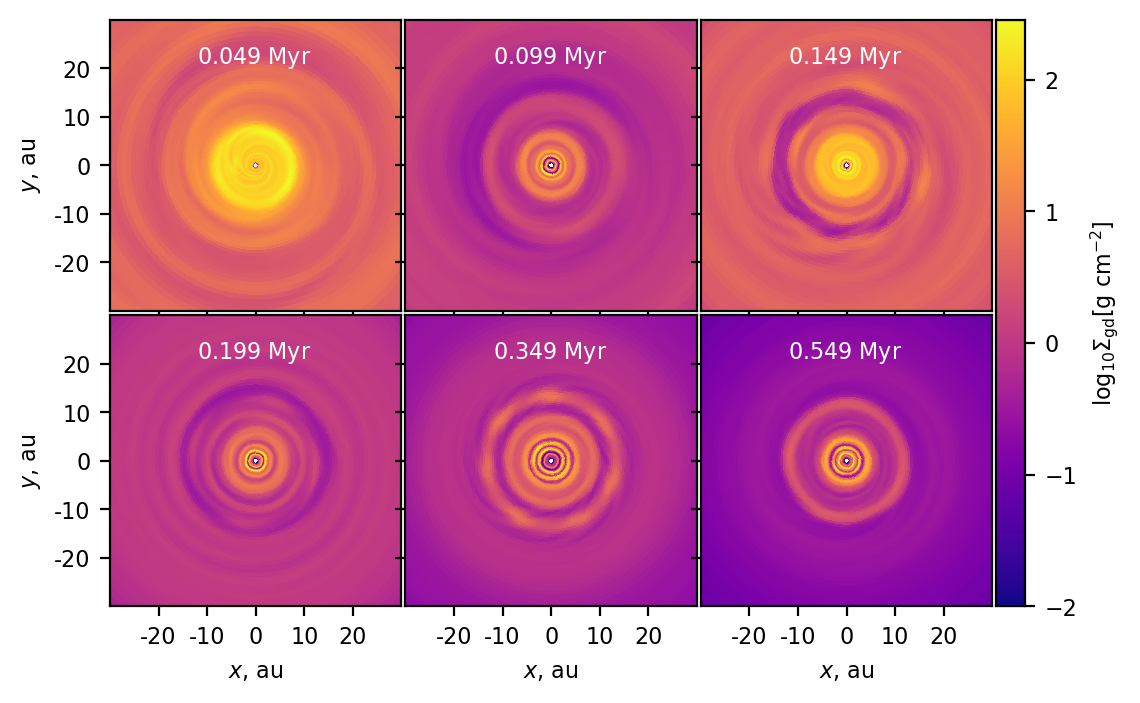}
\par\end{centering}
\caption{Two-dimensional distribution of the grown dust surface density in the inner $60\times 60$~au$^2$ box at time instances $0.049$, $0.099$, $0.149$, $0.199$, $0.349$, $0.549$~Myr after disk formation.  The time is counted from the disk formation instance. }
\label{fig:rings}
\end{figure}

Mass-loading from the envelope and transport of mass due to
viscous torques (they can be positive in the disk outer regions) lead to the disk spreading with time, as is evident from the black solid lines in Figure~\ref{Fig:radial}. A peak in the gas surface density develops at several au, while the outer parts are characterized by a 
power-law density profile, $\langle \Sigma_{\rm g}\rangle \propto r^{-1.5}$. The temperature decreases from $\simeq 700$~K at $r=1$~au to $20$~K at $r>1000$~au. The slope of the temperature profile is nearly $-0.5$ in the region $r<10$~au. Both density and temperature profiles exhibit a non-monotonic behaviour in the region $r<20$~au: there are variations of the surface density with amplitude of $\sim 1000-5000\,\mathrm{g}\,\rm{cm}^{-2}$ and temperature variations with amplitudes of $\sim 100-300$~K on spatial scales of $\sim 1$~au. These variations lead to the development of local pressure maxima and to the formation of dust rings discussed later in Sect.~\ref{longdisk}.
The dynamics of the disk is characterized by a slowly variable radial transport 
with $\langle v_r \rangle \le 0.01\,\rm{km}\,\rm{s}^{-1}$
within the disk, $r<200$~au, and fast infall onto the outer edge of the disk with velocity $\langle v_r \rangle \sim -0.1 \,\mathrm{km}\,\mathrm{s}^{-1}$. The radial profile of azimuthal velocity  follows a power-law dependence $\langle v_{\varphi} \rangle
\propto r^{-0.45}$ at $r<100$~au, i.e. the disk is slightly super-Keplerian. The $\langle 
v_{\varphi}\rangle$ radial profile is super-Keplerian  due to the contribution of disk 
self-gravity to the gravity of the central protostar.

\subsection{Long-term disk evolution}
\label{longdisk}
In Figure~\ref{Fig:S2D} we plot the spatial distributions of the gas surface density in the  $600\times 600$~au$^2$ box at different time instances after disk formation. The disk forms at $t\approx 0.056$~Myr after the onset of the collapse and its subsequent evolution can be divided into two phases. In the first phase, $t< 0.2$~Myr, the disk grows with time due to viscous spreading and continual mass-loading from the envelope. To illustrate the disk growth with time, we adopt $\Sigma = 1\,\mathrm{g \,cm}^{-2}$ as a typical surface density at the outer boundary of the disk. Figure~\ref{Fig:S2D} shows that the disk quickly grows up to a typical size of $100-150$~au by the time $t=0.2$~Myr. Spiral arms form at this time period, which reflect the development of gravitational instability in the disk. In the second phase, $t>0.2$~Myr, the disk radius only slowly increases with time. No spiral arms are observed in this  phase. Instead, ring structures form in the innermost region of the disk at $r\le 15-20$~au. These rings with a gas surface density of $\ga 200\,\mathrm{g \,cm}^{-2}$ persist till the end of the simulation. 
Similar ring-like structures were also reported in numerical hydrodynamics simulations of disk formation with an adaptive $\alpha$ by \citet{Kadam2019}.

These gas rings also represent the local pressure maxima, which help collecting grown dust into similar ring-like structure but of much higher contrast. Figure~\ref{fig:rings} presents the spatial distribution of grown dust surface density in the inner $60\times60$~au$^2$ box at the same evolutionary times as in Figure~\ref{Fig:S2D}. A series of concentric dust rings is formed in the inner ten astronomical units. The localization of the rings   coincides with the radial extent of the dead zone, as will be shown later in Figure~\ref{Fig:bursts}. 
{We note that turbulent dust diffusion (neglected in this study) may smooth out the dust rings obtained in our study, but low $\alpha$-values in the rings could require Schmidt numbers $\ll 1.0$ for this effect to be pronounced.  } Interestingly, the rings evolve with the disk, their radial position and shape change with time.
The total dust to gas mass ratio in the rings can reach 0.1, while in the rest of the disk it is reduced below the initial 1:100 value. The concentration of dust in the dead zones was also predicted in previous studies of protoplanetary disks  \citep[e.g.,][]{Dzyurkevich2010, Vorobyov2018, Kadam2019}.  Variations in the gas pressure and its radial gradient within the dead zone extent may be responsible for the formation
of a series of dust rings in our models. {Strong concentration of grown dust in the rings may be offset by including addition physics, such as turbulent diffusion or dependence of dust sticking properties on presence/absence of ices, which would also affect the level of dust backreaction on gas.} We postpone a more detailed analysis of the ring formation mechanisms, ring properties, and other implications of the dead zone (e.g., the Rossby wave instability) to a follow up study. 

\begin{figure}
\begin{centering}
\includegraphics[width=1\columnwidth]{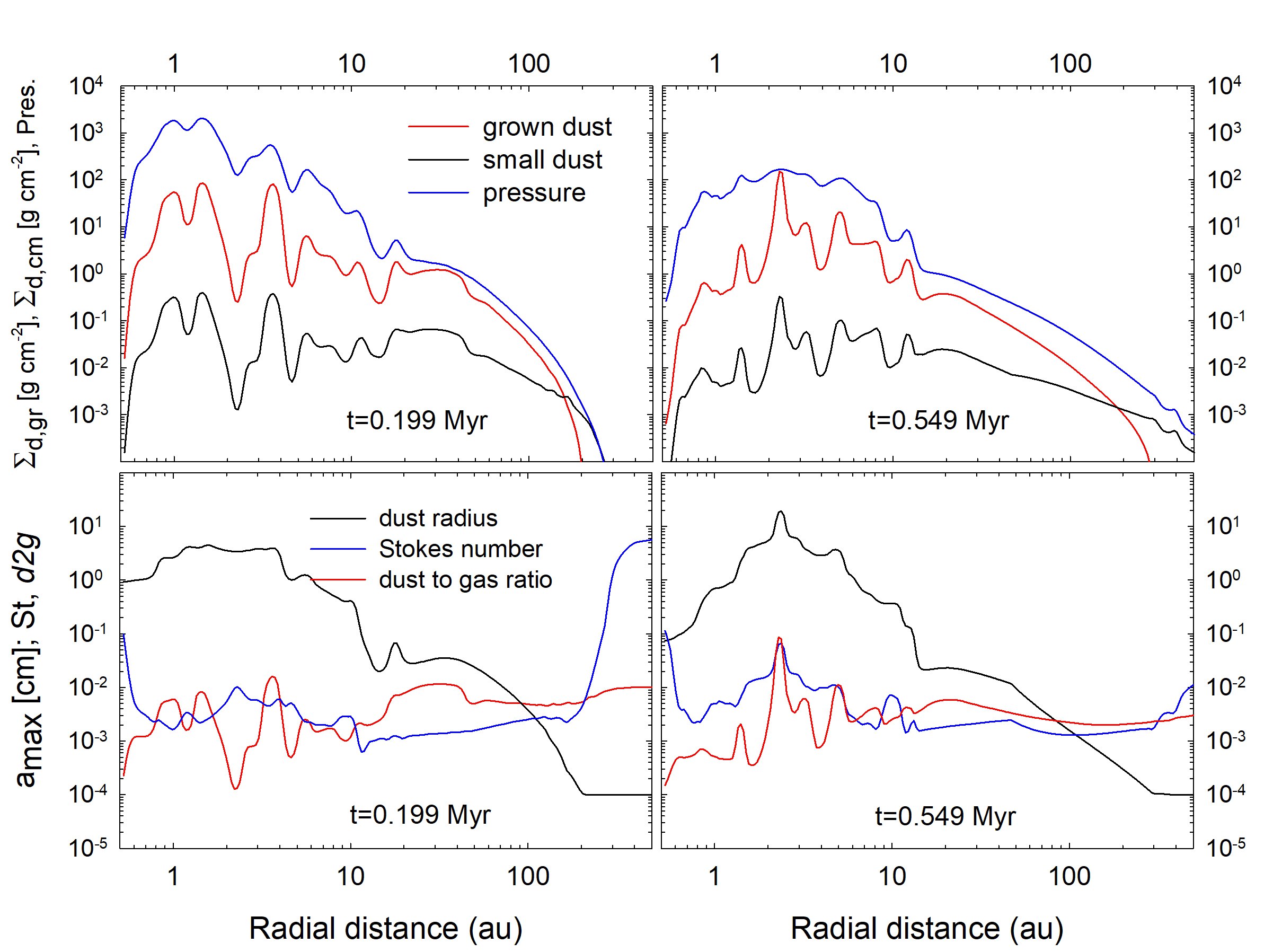}
\par\end{centering}
\caption{Azimuthally averaged dust characteristics at $t=0.199$~Myr (left column) and $t=0.549$~Myr (right column) from the instance of disk formation. The top row presents the surface densities of small and grown dust and also the vertically integrated gas pressure. The bottom row shows the dust radius, Stokes number, and total dust-to-gas ratio. }
\label{fig:azim}
\end{figure}

\begin{figure*}
\begin{center}
\includegraphics[width=0.48\textwidth, trim=0cm 0cm 0cm 0cm]{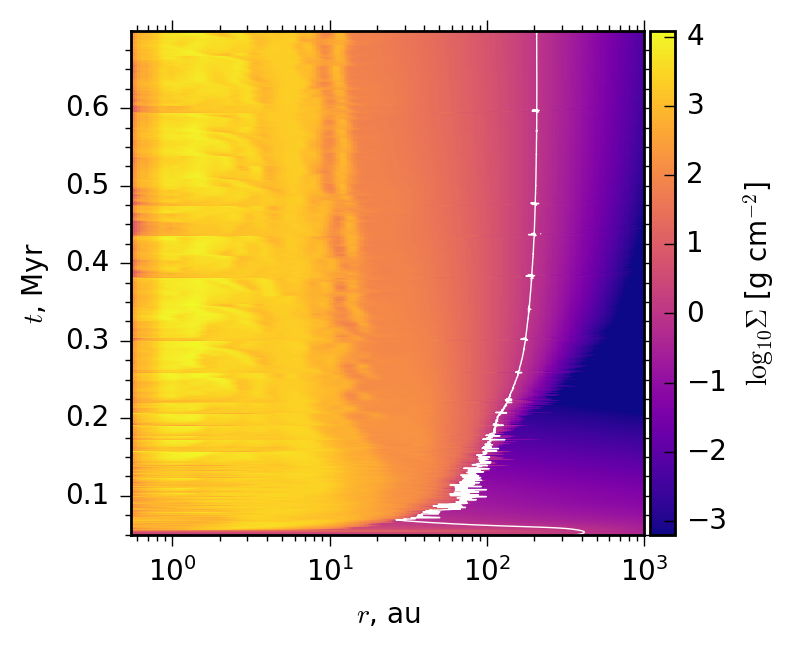}
\includegraphics[width=0.48\textwidth, trim=0cm 0cm 0cm 0cm]{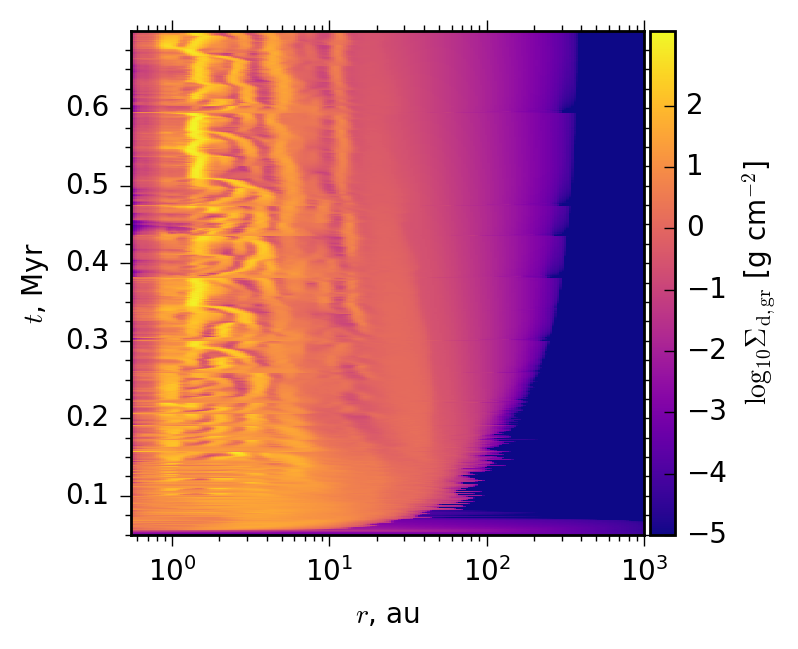}
\caption[width=0.68\textwidth]{Space-time diagrams of the azimuthally averaged radial profiles for the gas and grown dust surface densities (left and right panel, respectively). The white contour line indicates the disk loci with $\Sigma_{\rm g}=1.0$~g~cm$^{-2}$. }
\label{Fig:S_Sg}
\end{center}
\end{figure*}

Figure~\ref{fig:azim} presents the azimuthally averaged profiles of several quantities that describe in more detail the dust distribution in the disk. Two time instances corresponding to the bottom-left and bottom-right panels of Figure~\ref{Fig:S2D} are shown. The top row presents the surface densities of small and grown dust ($\Sigma_{\rm d,sm}$ and $\Sigma_{\rm d,gr}$, respectively) and also the vertically integrated gas pressure $\cal P$, while the bottom row shows the dust maximum radius $a_{\rm max}$, Stokes number $St=\Omega t_{\rm st}$, and total dust-to-gas mass ratio $d2g=(\Sigma_{\rm d,sm}+\Sigma_{\rm d,gr})/\Sigma_{\rm g}$. Here, $\Omega$ is the local angular velocity and $t_{\rm st}$ is the stopping time.
We note that we use the Keplerian angular velocity as a proxy for $\Omega$. This could underestimate the Stokes number because the disk gravity makes the disk rotate faster than what is expected from the pure Keplerian rotation.

Several interesting features  can be noted in the dust distribution. First, grown dust dominates over small dust in the bulk of the disk, except for the outer regions beyond 100~au. To remind the reader, all dust was initially in the form of small dust grains when the gravitational collapse commenced. The efficient conversion of small to grown dust is the result of dust growth, as can be seen from the black line in the bottom row. The maximum dust radius reaches 5.0~cm at $t=0.199$~Myr and 10~cm at $t=0.549$~Myr. The most efficient dust growth occurs in the inner 10~au where the dead zone is localized. Beyond the dead zone, $a_{\rm max}$ does not exceed a~few~$\times 10^{-2}$~cm. 
Second, the local peaks in the dust density coincide with the local pressure maxima, corroborating our previous suggestions about dust drift. Third, the dust-to-gas ratio is generally below 1:100, which was the initial value for the prestellar core. The exception are the most pronounced dust rings where $d2g$ can exceed 1:100 and reach 1:10 in the late disk evolution. The general decrease of $d2g$ throughout the disk is likely due to dust loss to the star during accretion bursts. Dust accumulates in the dead zone and is dumped on the star during bursts in higher amounts than would be expected without accumulation.  We checked the dust-to-gas mass ratio in the star and it is indeed slightly elevated 0.012.
Finally, the Stokes number is relatively low, hardly exceeding $10^{-1}$ within 100~au. A sharp increase of $St$ beyond 100~au is the result of a sharp increase in $t_{\rm st}$ in the rarefied outer regions. Our simulations show that a Stokes number on the order of unity, sometimes taken to be representative for protoplanetary disks \citep[e.g.][]{Booth2016}, may be an overestimate.

In Figure~\ref{Fig:S_Sg} we plot the space-time diagrams of the azimuthally averaged surface density profiles for gas, $\langle\Sigma_{\rm g}\rangle$, and grown dust, $\langle\Sigma\rangle_{{\rm d,gr}}$. Firstly, we consider the gas surface density distribution shown in the left panel.   The white contour line showing the locus of $\langle \Sigma_{\rm g}\rangle =1.0$~g~cm$^{-2}$ indicates that the disk quickly grows in time in the initial formation stage and reaches a radius of approximately 200~au by the end of the simulations.
Moreover, $\langle \Sigma_{\rm g}\rangle$ is highest in the inner several tens of astronomical units, but its radial distribution is not monotonic. Several rings of enhanced gas density appear in this region after 0.2~Myr and persist till the end of the simulations. The farthest ring is located at 15~au from the star.  Note the horizontal stripes that are evident in the left panel. These structures reflect MRI-triggered bursts that will be discussed in more detail in Sect.~\ref{Sec:Ideal_burst}.
The grown dust distribution shown in the right panel is more patchy. It also shows rings, but of higher contrast than those of the gas, reflecting efficient dust drift to the local pressure maxima (see also Fig.~\ref{fig:rings}). The grown dust distribution is also characterized by horizontal stripes akin to those seen in the left panel, but of higher contrast.


In Figure~\ref{Fig:T_B} we plot the space–time diagrams of the azimuthally averaged temperature and magnetic field profiles. The temperature inside the disk reaches values above $10^3$~K in the inner several astronomical units. Horizontal stripes akin to those of the gas and dust density distributions are also evident in the gas temperature distribution. Overall, the disk cools as it evolves. The distribution of $\langle B_z\rangle$ follows that of the gas surface density distribution, as $B_z\propto \Sigma_{\rm g}$ in the ideal MHD limit.

\begin{figure*}[t]
\begin{center}
\includegraphics[width=0.48\textwidth, trim=0cm 0cm 0cm 0cm]{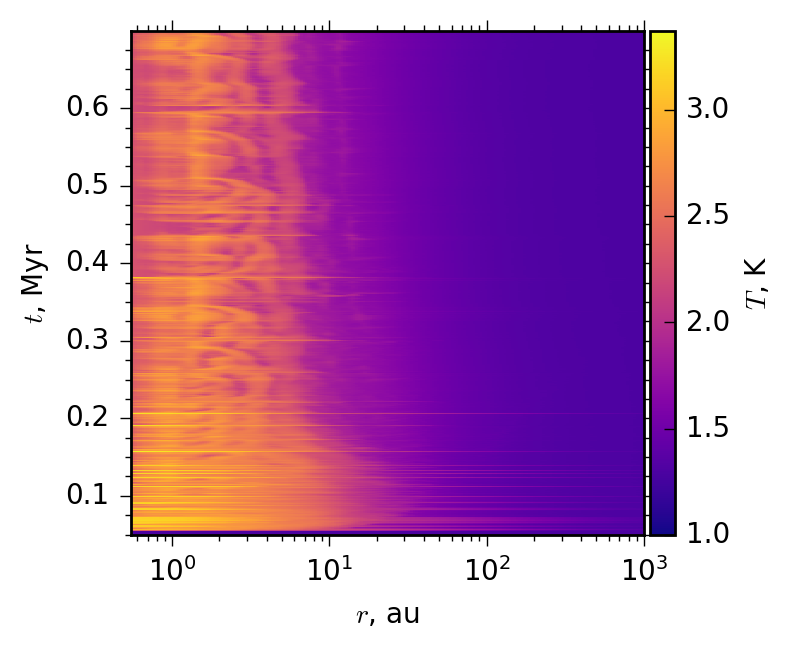}
\includegraphics[width=0.48\textwidth, trim=0cm 0cm 0cm 0cm]{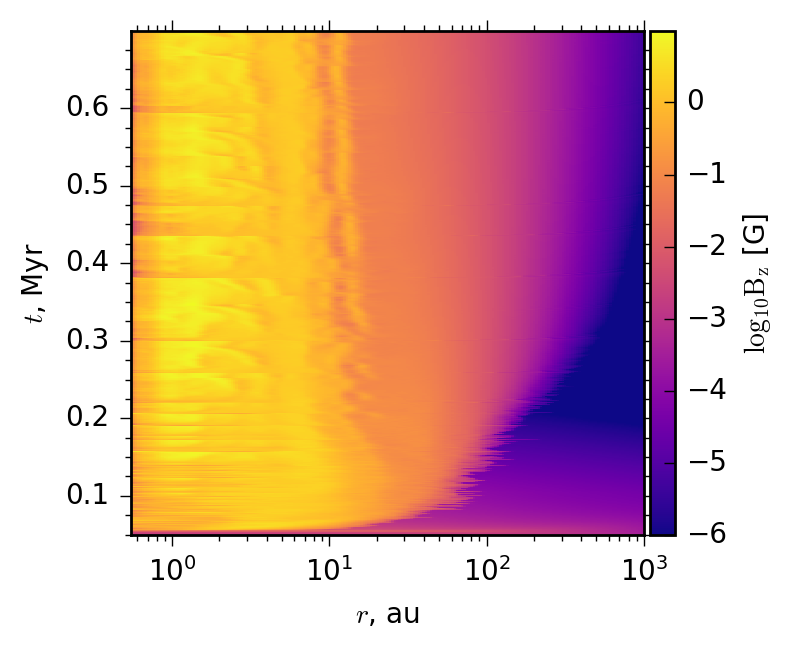}
\caption[width=0.68\textwidth]{Space-time diagrams of the azimuthally averaged radial profiles of temperature and $B_z$  (left and right panel, respectively).}
\label{Fig:T_B}
\end{center}
\end{figure*}


\section{MRI-triggered accretion bursts}

\label{Sec:Ideal_burst}

In this section, we analyse the temporal evolution of the rate of gas mass transport through the sink cell, $\dot{M}_{\rm g}=-2\pi r_{\rm sc} \Sigma_{\rm g} v_r$. We use this quantity as a proxy for the mass accretion rate on the star. In Figure~\ref{Fig:bursts}(a) we plot the time dependence of $\dot{M}$ starting from the disk formation instance till 0.7~Myr. The accretion rate is a few  $\times 10^{-6}\,M_{\odot}\,\mathrm{yr}^{-1}$ right after disk formation and in general decreases afterwards. This trend is accompanied by short-time variability, during which the accretion rate rapidly increases by 1--3 orders of magnitude. The time scale of variability is initially very short, so that $\dot{M_{\rm g}}(t)$  appears as a black strip in Figure~\ref{Fig:bursts}(a) at $t<0.1$~Myr. The amplitude of variability is initially limited by one order of magnitude,  but quickly increases to 2--3 orders of magnitude (see also Fig.~\ref{fig:1}).
In this early stage, the inner disk is sufficiently hot to support thermal ionization of alkaline metals and the disk is mostly in the MRI-active state. 

As the evolution continues and the inner disk cools, individual bursts of higher amplitude emerge as spikes in the $\dot{M_{\rm g}}(t)$ dependence after $t=0.1$~Myr. Six accretion bursts occur in a time interval from $0.2$~Myr after disk formation till 0.7~Myr. Figure~\ref{Fig:bursts}(b) shows part of the $\dot{M}(t)$ dependence for a time interval from $0.149$ to $0.150$~Myr. One accretion burst is observed in this period. It has a sharp rise, typical for many classical FUors (e.g., FU~Orionis and V1057), followed by a slow decline, and a sharp fall off after about 200~yr from the onset of the burst to a negligible value (implying that the matter stopped flowing towards the star for a short while). A small scale variability is evident during the outburst phase, which is also observed in some FUors \citep{Kospal2020}. 
We defer a more detailed analysis of the burst statistics  (peak luminosities, durations, rise times) to a follow-up paper.

\begin{figure*}[t]
\begin{center}
\includegraphics[width=0.98\textwidth, trim=0cm 0cm 0cm 0cm]{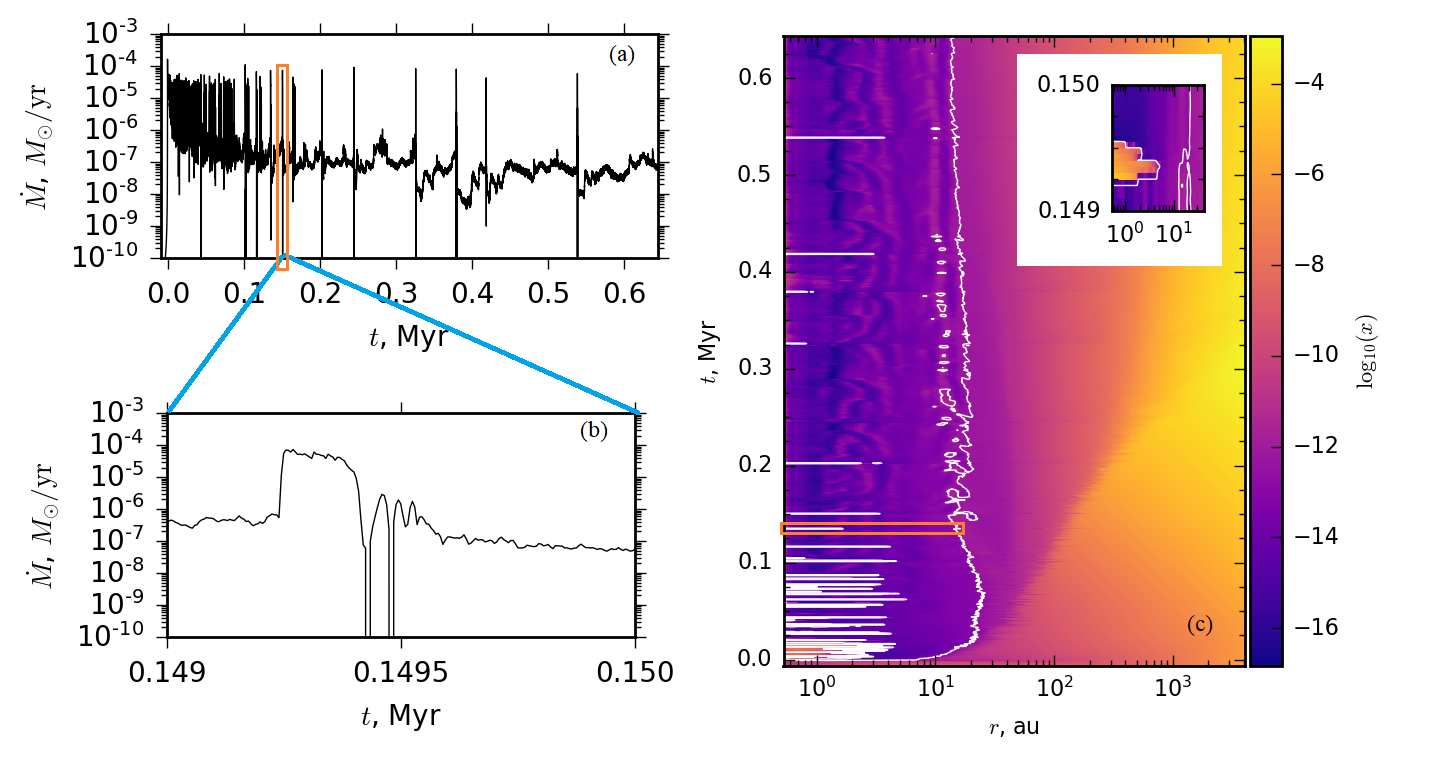}
\caption[width=0.68\textwidth]{ Top left (a): The accretion rate onto the star versus time starting from disk formation. Bottom left (b): similar to the top-left panel, but for a time interval $t=[0.149,\,0.150]$~Myr. Orange rectangles mark the burst analyzed in Figure~\ref{Fig:burst_038}.  Right (c): space-time diagram of the azimuthally averaged ionization fraction. The white contour line delineates the dead zone boundary. The time is counted from the instance of disk formation.}
\label{Fig:bursts}
\end{center}
\end{figure*}

To analyse the nature of the bursts, we plot the space-time diagram of the azimuthally-averaged ionization fraction $\langle x\rangle$ in Figure~\ref{Fig:bursts}(c). We also depicted the boundary of the dead zone with the white contour line in this plot. The dead zone is determined as the disk regions where $\Sigma_{\rm g}< \Sigma_{\rm crit}$ and the boundary of the dead zone is located at the disk loci where $\Sigma_{\rm g}=\Sigma_{\rm crit}$ (see Eq.~\ref{Eq:DZ}). The orange rectangles in all panels and the inset in Figure~\ref{Fig:bursts}(c) highlight the accretion burst occurring at $t\approx 0.1493$~Myr after the disk formation.
Figure~\ref{Fig:bursts}(c) shows that the ionization fraction is generally an increasing function of radius everywhere in the disk, except the innermost region. The smallest values of $x\sim 10^{-17}-10^{-16}$ are observed  in the region from $1$ to $10$~au. 
Comparison of Figures~\ref{Fig:S_Sg} and Figure~\ref{Fig:bursts}(c) reveals that the ionization fraction is lowest in the regions with highest gas and grown dust surface densities. This is because the rate of recombination onto the dust grains increases with gas density. Moreover, the ionization rate by cosmic rays decreases with increasing gas density due to cosmic ray attenuation.

The region of lowest ionization fraction is typically determined as a dead zone if $x\lesssim 10^{-13}-10^{-12}$ (see, e.g., Sano, 2000; DKH2014). In the considered case, the outer boundary of the dead zone lies at $r\approx 15-20$~au, just outside the dust rings in Figure~\ref{fig:rings}. Its position does not change appreciably during the disk evolution, except for the very early stages of disk formation. On the contrary, the location of the inner boundary of the dead zone is highly variable, mirroring the accretion variability in Figure~\ref{Fig:bursts}(a). For instance, white stripes in  Figure~\ref{Fig:bursts}(c) at $t<0.1$~Myr appear because the inner boundary of the dead zone frequently moves from $0.4$~au (inner boundary of the disk) to $r\approx 3-5$~au and back, reflecting similar high-frequency variations in $\dot{M_{\rm g}}$. 
The horizontal white stripes in $\langle x \rangle$ become less frequent with time and so do the mass accretion bursts in the top-left panel. These white stripes encompass time instances when the disk region at $r<5$~au becomes MRI-active and then MRI-dead again. The inset in Figure~\ref{Fig:bursts}(c) shows such an event occurred at a time interval from $0.149$~Myr to $0.150$~Myr. The lifetime of this MRI-active region is on the order of $200$~yr.

Figures~\ref{Fig:bursts} suggests that the accretion bursts occur when the inner region of the disk, $r<5$~au, becomes MRI-active. 
We refer to these events in the following text as the MRI-triggered bursts. To clearly demonstrate the causal link between the bursts, on the one hand, and ionization fraction, $\alpha$-value and gas temperature, on the other hand, we plot the $\dot{M}_{\rm g}$ dependence in the upper panel of Figure~\ref{Fig:burst_038} for a time interval from 0.149 to 0.150~Myr. Blue circles labelled as A, B and C mark the pre-burst, mid-burst, and post-burst stages, respectively. Additionally,  the two-dimensional spatial distributions of the ionization fraction (first row), $\alpha$-value (second row), and gas temperature (third row) at time instances corresponding to stages A, B, and C are plotted in the right panel of Figure~\ref{Fig:burst_038}. White contours in the two-dimensional panels delineate the dead zone boundary. Note that the inner boundary of the dead zone often coincides with the disk inner edge and is hence not visible in these plots.

Figure~\ref{Fig:burst_038} shows that the ionization fraction is lowest ($\approx 10^{-14}$) near the inner disk boundary and the dead zone extends to $r\approx20$~au in the pre-burst stage~A. We note that the dead zone is not continuous and may be patchy, reflecting the ring-like structure of the inner disk.
The ionization fraction is on the order of $10^{-13}$ at the boundary of the dead zone.
The $\alpha$-value in the dead zone is lower than 0.01 and the gas temperature does not exceed a few hundred Kelvin.
At $t=0.1493$~Myr (state B), the ionization fraction becomes higher, $x\sim 10^{-8}-10^{-7}$, in the innermost region of the disk, $r<5$~au. The $\alpha$-value and gas temperature also jump to 0.1 and $>1000$~K, respectively.  This region becomes MRI-active, which leads to an enhanced accretion rate. After accretion burst, the active region at $r<5$~au disappears and the accretion rate falls down to a pre-burst value (stage C). The $\alpha$-value, and ionization fraction return to their pre-burst values, while the inner disk gradually cools down to the pre-burst state.

\begin{figure*}
\begin{center}
\includegraphics[width=0.98\textwidth, trim=0cm 0cm 0cm 0cm]{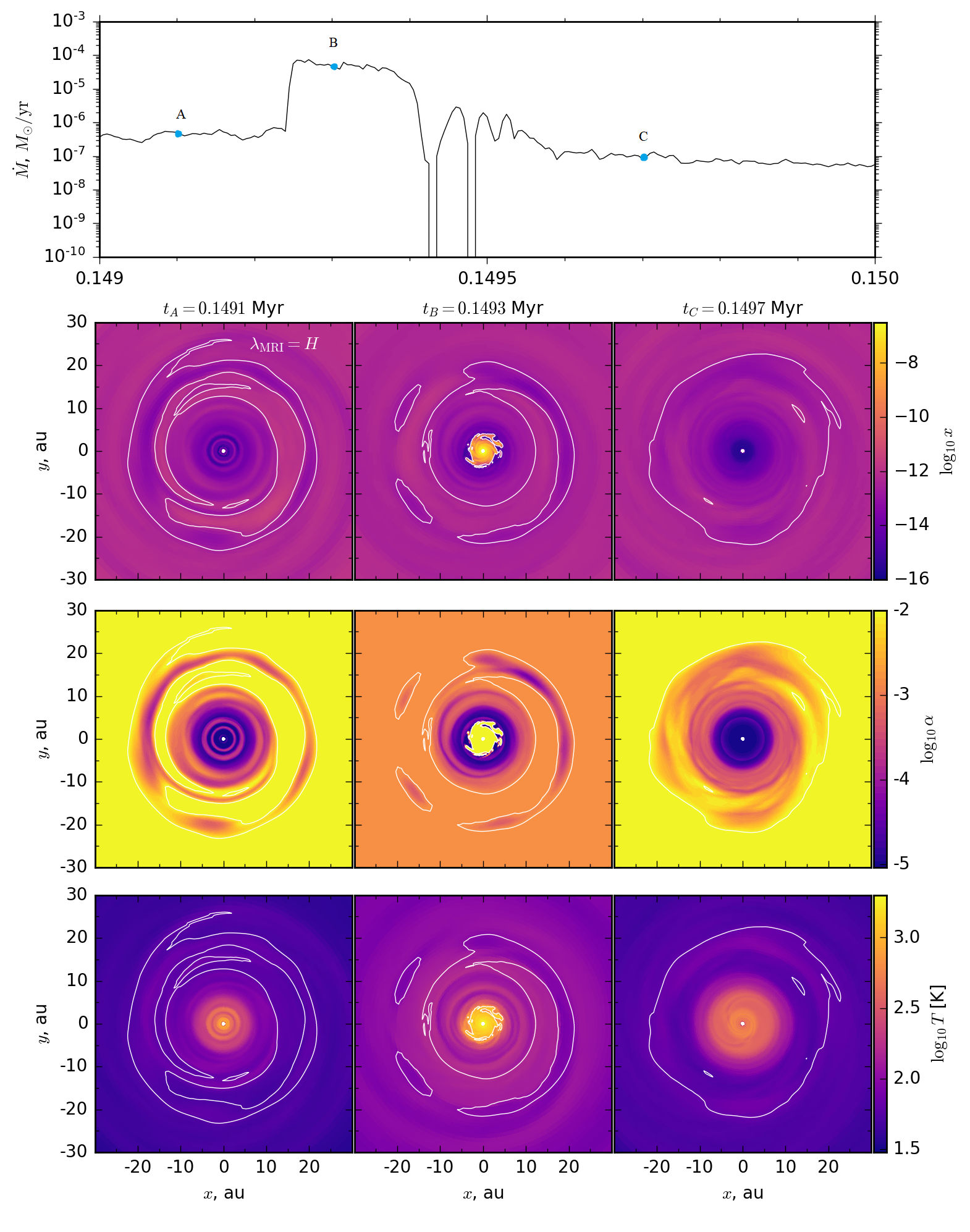}
\caption[width=0.68\textwidth]{Upper panel: accretion rate onto the star versus time in the time interval $t=[0.149,\,0.150]$~Myr after the disk formation. Lower rows of panels: two-dimensional distributions of the ionization fraction (first row), turbulent parameter $\alpha$ (second row), and gas temperature (third row) in the region $r\leq 30$~au at time instants $t=0.1491$, $0.1493$ and $0.1497$~Myr marked with blue rounds with labels 'A', 'B' and 'C' in the upper panel, respectively. The white contour delineates the boundary of the dead zone.}
\label{Fig:burst_038}
\end{center}
\end{figure*}

To further analyze the burst mechanism, we plot in Figure~\ref{fig:burst_time} the time evolution of several quantities at the sink--disk interface (0.52~au) as the burst emerges and decays. In particular, the top row shows the mass accretion rate $\dot{M}_{\rm g}$ and total luminosity $L_{\rm tot}$, while the other panels show the midplane temperature $T_{\rm mp}$, gas and grown dust surface densities $\Sigma_{\rm g}$ and $\Sigma_{\rm d,gr}$, $\alpha$-parameter, and ionization fraction $x$ at the sink-disk interface (0.52~au).  The pre- and post-burst total stellar luminosity is about 3~$L_\odot$ and the peak luminosity reaches 165~$L_\odot$. Small-scale variability during the active phase and also some moderate after-burst variability is also present.  The midplane temperature before the burst is  650--700~K and the ionization fraction is below $10^{-13}$. However, $T_{\rm mp}$ rises gradually due to heating provided by residual viscosity with $\alpha=10^{-5}$ and adiabatic compression provided by inflowing matter and spiral arms \citep{Bae2014}  until thermal collisions begin to ionize alkaline metals. This is a runaway process that depends exponentially on the temperature (see Eq.~\ref{x_thermal}). The rise in ionization is accompanied by the rise in viscous heating (via increased $\alpha$), causing temperature to increase further and finally causing the burst. We note that the concentration of dust in the dead zone assists the process of MRI triggering because the increased optical depth makes the dead zone easier to warm up.
We warn, however, that we use dust opacities that do not depend on dust growth \citep{Semenov2003} and more accurate computations with dust-growth-dependent opacities are needed to confirm the effect.

At the onset of the burst $T_{\rm mp}$ jumps sharply to 1800~K, but quickly declines below the pre-burst value as the burst develops and decays. The sharp rise in the midplane temperature is caused by strong viscous heating when the dead zone turns into an MRI-active region with a high $\alpha$-value and drop in $T_{\rm mp}$ is caused by radiative cooling as the inner disk losses mass and becomes less optically thick to its own thermal radiation.
The surface densities of gas and grown dust also drop during the burst, reflecting the loss of disk mass during the burst. We note that the drop in $\Sigma_{\rm d,gr}$ is deeper than in $\Sigma_{\rm g}$, which can be explained by efficient accumulation and drift of grown dust in the dead zone prior the burst. As is expected, the $\alpha$-value increases sharply during the burst and drops back to the pre-burst value when the burst is over. The same is true for the ionization fraction.
It is worth noticing that both the mass accretion rate and total luminosity return to almost pre-burst values earlier than the burst is formally over, as is signalized by the drop of $\alpha$ and $x$ to the pre-burst state.

\begin{figure}
\begin{centering}
\includegraphics[width=1\columnwidth]{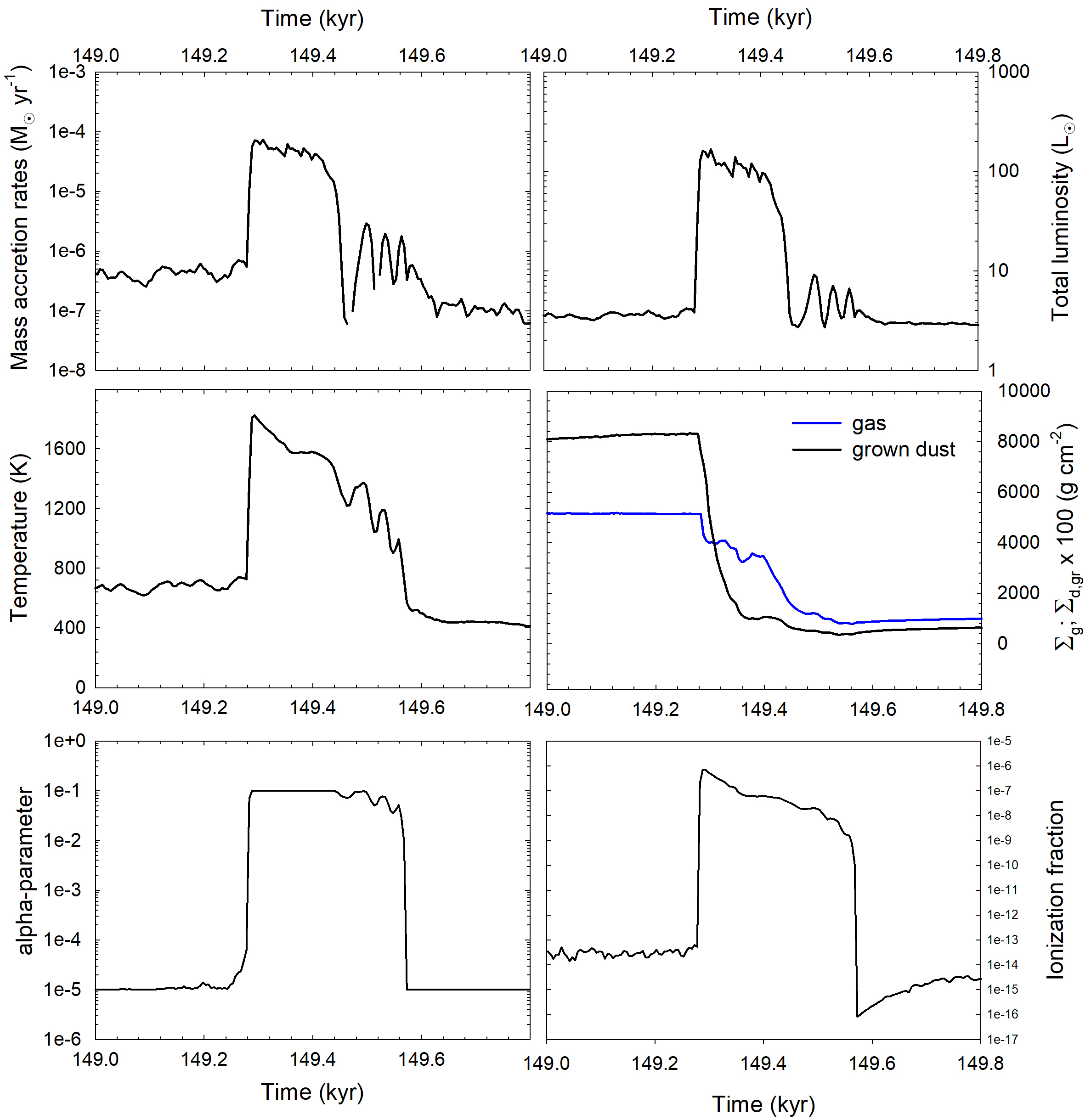}
\par\end{centering}
\caption{\label{fig:burst_time} Time evolution of several disk and burst characteristics at the sink--disk interface as the burst develops and decays. Shown are the mass accretion rate (top-left), total luminosity (top-right), midplane temperature (middle-left), gas and grown dust surface densities (middle-right), $\alpha$-value (bottom-left), and ionization fraction (bottom-right). }
\end{figure}


\section{Parameter space study}
\label{ParamStudy}

\begin{figure}
\begin{centering}
\includegraphics[width=1\columnwidth]{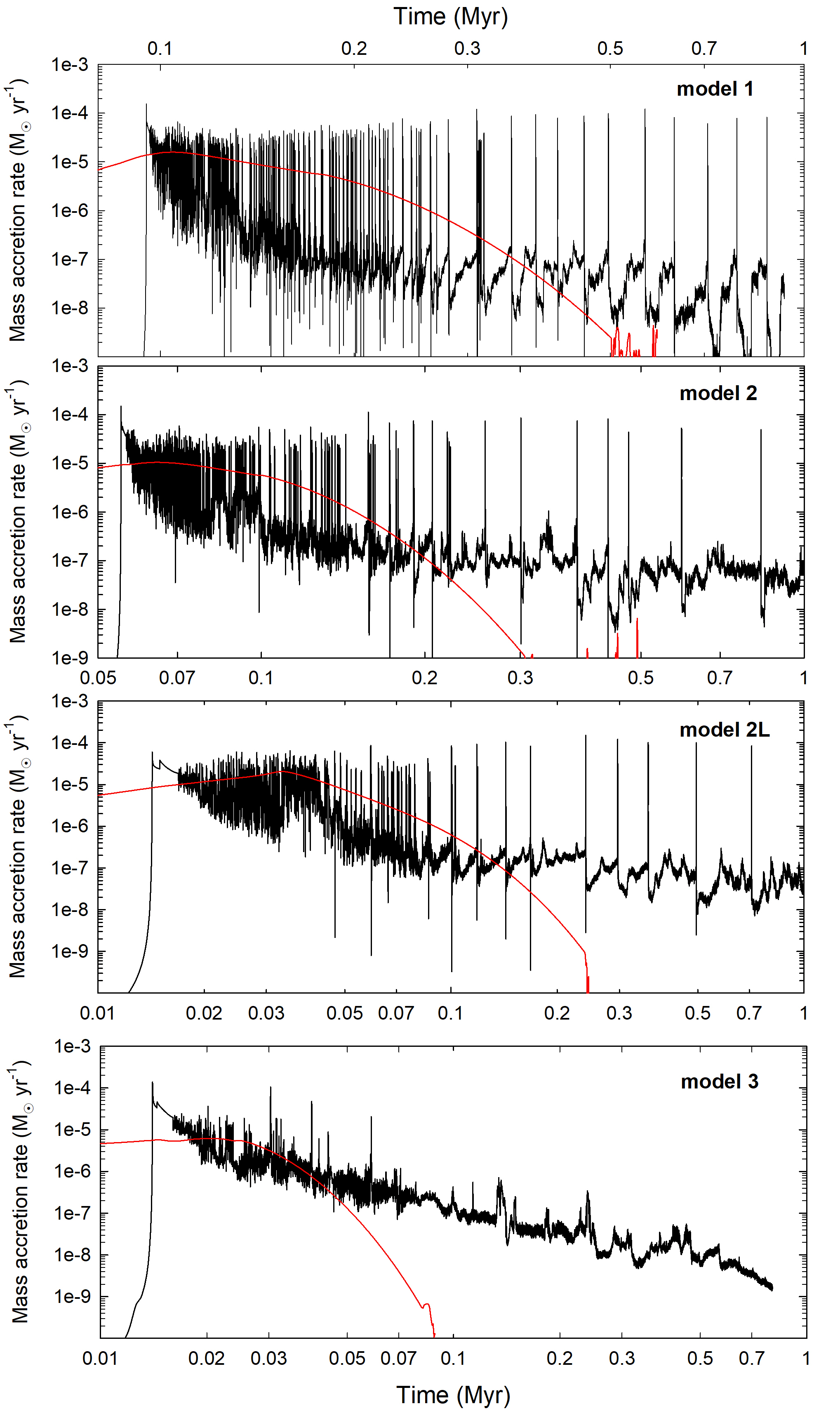}
\par\end{centering}
\caption{\label{fig:1} Protostellar accretion rates (black lines) and envelope infall rates (red lines) as a function of time in models with various initial pre-stellar come masses and mass-to-flux ratios: model~1 -- $M_{\rm core}=1.45~M_\odot$ and $\lambda=2$, model~2 -- $M_{\rm core}=0.83~M_\odot$ and $\lambda=2$, model~3 -- $M_{\rm core}=0.21~M_\odot$ and $\lambda=2$, and model~2L -- $M_{\rm core}=0.83~M_\odot$ and $\lambda=10$. The time is shown on the log scale to better resolve the bursts in the early evolution stage. }
\end{figure}

Unlike many previous studies of the burst phenomenon, we can consider the effects of different initial conditions in pre-stellar cores on the development of MRI-triggered bursts in protoplanetary disks. Here, we vary the initial pre-stellar core mass and strength of magnetic field. Figure~\ref{fig:1} presents the resulting mass accretion rates in four models, the parameters of which are listed in Table~\ref{table1}.
More specifically, models 1, 2, and 3 have progressively smaller masses of the pre-stellar core ($M_{\rm core}$), while model 2L is characterized by an increased mass-to-flux ratio $\lambda$.
The black lines show the mass accretion rate from the disk through the sink cell and the red lines correspond to the mass infall rate from the envelope onto the disk. The latter quantity is calculated as the mass flux at a radial distance of 1000~au from the star, except for model 3, in which case the infall rate is calculated at 400~au (this model is characterized by a compact initial core).

The accretion rate histories are different in models with distinct pre-stellar core masses. The higher mass models with $M_{\rm core}=1.45$ and 0.83~$M_\odot$ (first and second panels) exhibit multiple bursts throughout the entire considered period of evolution (1.0 Myr). In particular, $\dot{M}_{\rm g}$ in the early evolution is highly variable and the system spends roughly equal time in states of low- and high-rate accretion. The amplitude of variability spans one-to-two orders of magnitude. After about 0.08--0.1~Myr the accretion rate begins to clearly exhibit individual bursts interspersed with longer periods of low-rate quiescent accretion. This qualitative change in the behaviour of protostellar accretion coincides with the time when the envelope infall rate begins to decline, reflecting the gradual depletion of the envelope material via infall on the disk. The frequency of the bursts further reduces with declining mass infall rate. 

The apparent correlation of the burst activity with the disk mass-loading can be explained by the action of gravitational instability, which is strongest in the embedded phase thanks to mass-loading from the envelope that replenishes the disk mass lost via accretion onto the star. The associated gravitational torques are usually stronger than the viscous torques in the embedded phase \citep{VB2009}, so that the both torques are 
very efficient in supplying the inner disk with matter and recharging the inner disk for the MRI-bursts to occur. When the envelope depletes, the gravitational instability diminishes and only the viscous torques remain operational, resulting in longer recharge times and longer quiescent periods. A similar phenomenon was described in
\citet{Zhu2009a}.

\begin{figure*}
\begin{centering}
\includegraphics[trim= 0mm 0mm 2mm 0mm, width=0.75\columnwidth]{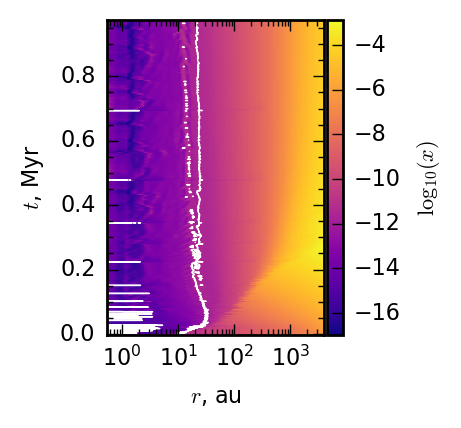}
\includegraphics[trim= 2mm 0mm 0mm 0mm, width=0.75\columnwidth]{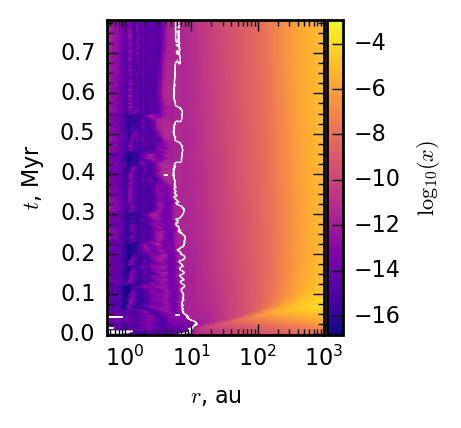}
\par\end{centering}
\caption{\label{fig:2vs2l} Space-time diagrams showing the ionization fraction in model~2L (left) and model~3 (right). The white contour lines delineate the position of the dead zone. }
\end{figure*}

The lower mass model with $M_{\rm core}=0.21~M_\odot$ and the terminal stellar mass $M_{\rm \ast,fin}=0.17~M_\odot$ (fourth panel) demonstrates
a different behavior. Accretion bursts occur only occasionally and are of much smaller amplitude than in the higher mass models. The mass accretion rate gradually declines with time to a small value of $10^{-9}~M_\odot$~yr$^{-1}$ by the end of the simulation, although order-of-magnitude variations can still be present at late evolutionary times. The mass infall rate declines much faster in this model, meaning that the embedded phase in this model is short-lived, less than 0.1~Myr.

Model 2L with an increased mass-to-flux ratio, but other characteristic similar to model~2, forms the disk notably earlier due to faster collapse of a less-magnetized core. The early disk evolution is characterized by a highly variable accretion  and well defined bursts start occurring 50~kyr after the onset of collapse. The number and frequency of the bursts are not strongly affected by an increase of the mass-to-flux ratio by a factor of 5.

To understand the difference in the burst behavior between the considered models, we plot in Figure~\ref{fig:2vs2l} the space-time diagrams showing the time evolution of the azimuthally averaged ionization fraction $\langle x \rangle$ in model~2L (left) and model~3 (right). A comparison with the right panel of Figure~\ref{Fig:bursts} showing the corresponding quantity for model~2 reveals that the dead zone has a different extent in the three models considered. It is largest in model~2L, with the outer boundary reaching 30~au in the early stages, and is smallest in model~3, extending only to 6--8~au. Consider first model~2L with an increased mass-to-flux ratio and reduced magnetic field. Equation~(\ref{Eq:DZ}) indicates that reducing $B_z$ also lowers the value of $\Sigma_{\rm crit}$ (in the limit of $\Sigma_{\rm crit} \rightarrow 0$ the entire disk is MRI-dead) and broadens the dead zone.  This trend is somewhat offset by a mild increase in the ionization fraction. This occurs because the model with reduced magnetic field forms a more massive and warmer disk (strong magnetic fields delay the disk formation, spreading its evolution over a longer time). The net effect is a mild broadening of the dead zone in model~2L as compared to model~2 with stronger magnetic support. 

In the case of model~3, the dead zone shrinks because the column density of gas decreases. This model has a much smaller reservoir of mass in the collapsing core and forms a disk of notably lower mass, $0.007~M_\odot$, in contrast to  $0.11~M_\odot$ in model~2. Both values are referred  to the end of simulations. The disk densities and temperatures are appreciably lower in model~3 and the thermal ionization of alkaline metals occurs only in the very early stages of disk evolution, when it is still relatively dense and warm thanks to mass-loading from the infalling envelope.
To summarize, it appears that there exists a minimal mass of prestellar cores and stellar masses, below which the MRI-bursts in subsequently formed disks are unlikely to occur. A similar conclusion was made in \citet{Kadam2020}. We checked if models 1, 2, and 2L may have bursts when the stellar mass is below $0.2~M_\odot$ (but grow later due to accretion). It turned out that when the clear-cut individual bursts start to occur (skipping the initial highly variable stage it is 0.11~Myr for model~1, 0.08~Myr for model~2, and 0.04~Myr for model 2L) the stellar mass is already above 0.25--0.3~$M_\odot$. Reducing the strength of magnetic field by a factor of 5 does not seem to have a notable effect on the frequency of accretion bursts. Finally, we note that our model does not reproduce the EX Lupi-like luminosity outbursts, which are known to occur in the Class II--III stages of disk evolution, most likely because these outbursts are caused by phenomena that operates much closer to the star than we can resolve in our global disk simulations \citep{Dangelo2012,Armitage2016}.


\section{Comparison with previous studies and model caveats}
\label{comparison}
The MRI-triggered accretion bursts were studied in detail in a series of papers \citep[e.g.,][]{Armitage2001,Zhu2009a,Zhu2009,Martin2012}, but here we focus on two most relevant studies by \citet{Bae2014} and \citet{Kadam2020}. Both studies employ numerical hydrodynamics simulations in the thin-disk limit and use an adaptive $\alpha$-parameter method to simulate the MRI bursts. The thermal physics is also similar, although \citet{Bae2014} used a more sophisticated two-temperature description of the disk vertical structure. The latter authors considered the effect of disk mass-loading from the envelope in a parametric manner, while \citet{Kadam2020} and we use a global model that computes the dynamics of both the disk and envelope in one simulation and on one numerical grid. The inner computational boundary in \citet{Bae2014} and \citet{Kadam2020} was set at 0.2~au and 0.41~au, respectively, while in our study it is set equal to 0.52~au. A larger inner boundary in our study and in \citet{Kadam2020} is dictated by a heavier numerical load on a finer numerical resolution ($256\times 256$ or $512 \times 512$) in contrast to $128 \times 128$ in \citet{Bae2014}.

The main difference of our work from aforementioned studies is that we take  magnetic fields explicitly into account (albeit in a simplified flux-freezing approximation), evolve dynamically both the gas and dust disk subsystems, and also calculate explicitly the ionization fraction. This allowed us to reduce the number of free parameters when calculating the adaptive $\alpha$-parameter (see Eq.~\ref{alphavalue}). For instance, \citet{Bae2014} and \citet{Kadam2020} used a critical temperature for MRI activation, $T_{\rm crit}$, set equal to 1300 to 1500~K. In our study this parameter is absent and the MRI activation depends on the explicitly calculated ionization fraction. Furthermore, \citet{Bae2014} and \citet{Kadam2020} used a fixed value for the MRI-active disk layer, $\Sigma_{\rm MRI}$, set equal to 10 or 100~g~cm$^{-2}$. In our approach, $\Sigma_{\rm MRI}$ is a time- and space-varying quantity calculated from the strength of magnetic field and ionization fraction (see Eq.~\ref{Eq:DZ}).

Our simulations can reproduce the main features of the MRI-triggered bursts reported in \citet{Bae2014}, especially when they take disk self-gravity into account and adopt a non-zero value for $a_{\rm dz}$. In particular, the dependence of the burst occurrence frequency on the evolutionary stage is well reproduced. On the other hand, the magnitude and the duration of the bursts seem to differ. For instance, our model predicts  shorter bursts  of higher magnitude (a few hundred years with $\dot{M}\approx 5\times 10^{-5} - 10^{-4}~M_\odot$~yr$^{-1}$), while \citet{Bae2014} obtained longer and less energetic bursts (a few thousand years with $\dot{M}\approx \mathrm{a~few}\times10^{-5}~M_\odot$~yr$^{-1}$). This can be explained by a higher value of the $\alpha$-parameter chosen to represent the MRI-active state during the outburst. In our model it is set equal to 0.1, while in \citet{Bae2014} it was 0.01. We cannot reliably compare the shape of the outbursts because we considered only one example in our study, but our considered burst (see Fig.~\ref{Fig:bursts}) is very similar to that shown in fig.~2 of \citet{Bae2014}, but is quite different from other cases presented in that study. A detailed comparison of the burst shapes is postponed for a follow-up study.

When compared to the work of \citet{Kadam2020}, our study can reproduce one of their key conclusions stating that there may exist a lower limit on the prestellar core mass $M_{\rm core}$ for the MRI bursts to occur. In terms of the final protostellar mass, \citet{Kadam2020} quoted $M_{\rm \ast,fin}=0.28~M_\odot$ as the lower limit, while in our study it is $0.17~M_\odot$ (see Table~\ref{table1}). This difference may partly be attributed to a coarse grid of models considered in both studies. \citet{Kadam2020} also considered different critical temperatures $T_{\rm crit}$ for the MRI activation and demonstrated that the burst activity reduces with increasing $T_{\rm crit}$. We do not have this parameter in our model, but the burst occurrence frequency better matches the models of \citet{Kadam2020} with $T_{\rm crit}=1300$~K.

The main result of our study is that the MRI-triggered bursts occur for a wide range of model realizations ($M_{\rm core}$, $\lambda$) and over long evolutionary times, meaning that this is a robust phenomenon. Nevertheless, several model caveats need to be mentioned.  We chose a fairly low mass-to-flux ratio $\lambda=2$, implying strong magnetic fields. The perturbation analysis indicates \citep[see, e.g.,][]{Armitage2015} that the MRI can be suppressed if magnetic fields are too strong. The instability criterion for the development of the MRI in terms of the plasma $\beta$-parameter reads
\begin{equation}
\beta \ge {2\pi^2 \over 3},
\label{MbetaArm}
\end{equation}
where $\beta=8 \pi P/B_z^2$ and $P$ is the gas pressure in the disk midplane. This criterion can be expressed in terms of the $\lambda$ ratio as
\begin{equation}
    \beta = C_0 \lambda^2 Q \ge {2\pi^2 \over 3},
\label{Mbeta}
\end{equation}
where $Q$ is the Toomre parameter of a Keplerian disk and $C_0$ is a constant on the order of unity depending on how we convert the volume density into the surface density\footnote{When deriving Eq.~(\ref{Mbeta}) we assumed that $H_{\rm g}=c_{\rm s}/\Omega$. In massive disks with $Q\approx 1$ a more accurate but complex expression should be used \citep[see eq. 11 in][]{Kratter2016}.}. For a Gaussian distribution in the vertical direction $C_0=\sqrt{2/\pi}$ and for a constant distribution $C_0=1$. From Equation~(\ref{Mbeta}) and for our choice of $\lambda=2$ it follows that the MRI can be suppressed if $Q<2$. In the region of interest in our study where the MRI bursts are triggered ($r\le 5$~au) the $Q$-parameter is always greater than 2 because these regions are fairly hot (see Fig.~\ref{Fig:T_B}). Therefore, Criterion~(\ref{MbetaArm}) is fulfilled in the innermost disk, but may break in more distant regions where $Q$ drops below 2 and the disk becomes gravitationally unstable. This occurs in the initial stages of disk evolution (see Fig.~\ref{Fig:S2D}). However, in these early stages the mass and angular momentum transport in the corresponding disk regions is dominated by gravitational torques, and not by viscous ones \citep{VB2009}, and the pumping rate of the MRI should not be strongly affected. Our analytical estimates are reinforced by the direct calculation of the azimuthally averaged $\beta$-parameter shown in Fig.~\ref{fig:beta} as a function of time for the fiducial model. The red line encompasses the disk regions with $\beta < 6.0$ where the MRI may not develop. Clearly, this region partly coincides with the already existing dead zone outlined by the white line and diminishes after 0.2~Myr. The innermost regions involved in the bursts are not affected.

\begin{figure}
\begin{centering}
\includegraphics[width=1\columnwidth]{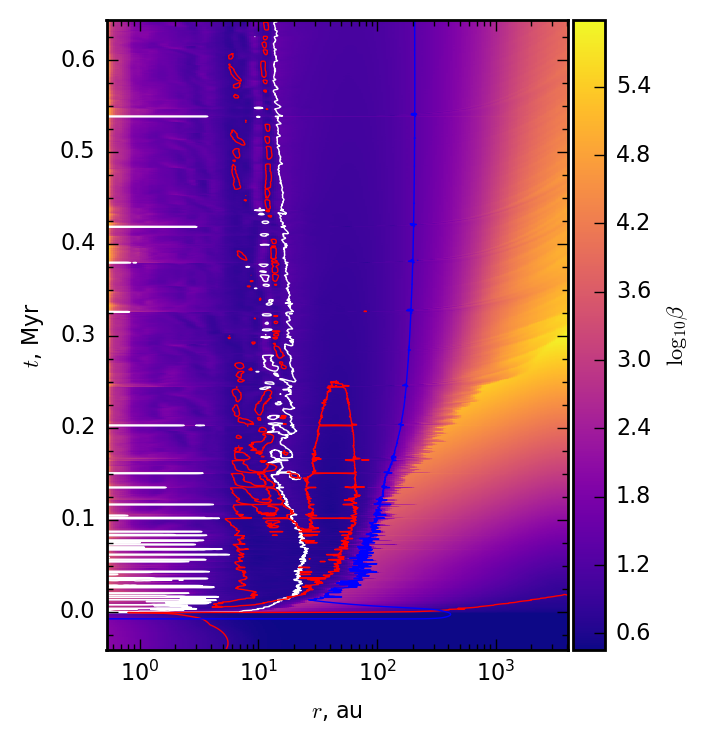}
\par\end{centering}
\caption{\label{fig:beta} Space-time diagram of the plasma $\beta$-parameter for the fiducial model. The red line outlines the regions with $\beta<6.0$. The white line shows the extent of the dead zone defined by Eq.~(\ref{Eq:DZ}). The blue line outlines the outer edge of the disk defined by $\Sigma_{\rm g}=1.0$~g~cm$^{-2}$. }
\end{figure}

We considered the ideal MHD limit and neglected the non-ideal MHD effects due to a heavy computational load. The diffusion of magnetic field in the inner disk regions may affect the frequency and the shape of the bursts, as Equation~(\ref{Eq:DZ}) implies. The size of the inner computational boundary may also influence the burst statistics and properties. The thermal instability may be activated if the sink cell is further reduced in size, as was demonstrated in \citet{Kadam2020}. Although this is the first study of the MRI-triggered burst phenomenon where dust dynamics and growth is explicitly taken into account, our model of dust growth is still simplistic and needs improvement, with the ultimate goal to employ the Smoluchowski-type integro-differential equation \citep{Drazkowska2019}.
We have not considered the effects of magnetocentrifugal disk winds that can be efficient in removing mass and angular momentum in the disk regions from several to several tens of astronomical units \citep[e.g.,][]{Suzuki2016,Bai2016,Guedel2019}, thus potentially affecting the burst activity by reducing the rate of mass accumulation in the innermost disk regions. It is, however, not easy to take disk winds in the thin-disk hydrodynamic models into account (which may require to relax the zero external current condition in our case) and we will consider several options in the future on how to proceed in this direction. Finally, we note that the effect of magnetic braking and a better calculation of the ionization fraction should be considered in a future development of the model.

\section{Conclusions}
\label{conclude}

We studied numerically the formation and long-term ($\sim 1.0$~Myr) evolution of protoplanetary disks starting from the gravitational collapse of prestellar cores with a  special emphasis on the development of MRI-triggered accretion bursts in the innermost disk regions. We employed magnetohydrodynamical simulations in the thin-disk limit adopting the ideal MHD approximation. The numerical model also features the co-evolution of gas and dust including dust backreaction on gas and dust growth. 

To simulate the MRI-triggered bursts, we employed the adaptive $\alpha$-parameter method of \citet{Bae2014}, upgraded to take the dependence of the MRI bursts on the magnetic field strength and ionization fraction explicitly into account. In this method, the turbulent $\alpha$-parameter is a weighted average between $\alpha_{\rm dz}=10^{-5}$, corresponding to the MRI-dead disk regions, and $\alpha_{\rm MRI}=10^{-2}$--$10^{-1}$, typical for the fully MRI-active disks.  Disk self-gravity was explicitly taken into account via the solution of the Poisson integral. 
Four models with different masses of prestellar cloud cores ($M_{\rm core}$) and  mass-to-magnetic flux ratios $\lambda$ were considered.
Our main results can be summarized as follows:

-- A dead zone in the inner disk region forms soon after disk formation, which is characterized by a low ionization fraction ($x \la 10^{-13}$) and temperature on the order of several hundred Kelvin.
The outer boundary of the dead zone extends from a few to several tens of astronomical units, depending on $M_{\rm core}$ and $\lambda$. The inner boundary coincides with the inner edge of the disk (imposed by the inner numerical boundary at 0.52~au). The dead zone features diffuse gas and sharp dust rings, the latter forming thanks to radial drift of dust towards the local pressure maxima, an effect also predicted in previous studies of dead zones in protoplanetary disks  \citep[e.g.,][]{Dzyurkevich2010, Vorobyov2018, Kadam2019}. In the rest of the disk, however, the dust-to-gas mass ratio drops lower than the initial 1:100 value.

-- Gradual warming of the dead zone (due to residual viscosity and compressional heating, see \citet{Bae2014}) leads to the thermal ionization of alkaline metals and a sharp increase in the ionization fraction. An accretion burst ensues as the inner several astronomical units of the disk become MRI-active and the turbulent viscosity sharply increases. The accretion rate onto the star rises to $5\times 10^{-5}$--$10^{-4}~M_\odot$~yr$^{-1}$ and the accretion luminosity exceeds $100~L_\odot$. The burst is characterized by a sharp rise, a plateau with short-term variability, and a fast decline as the inner disk becomes depleted of matter. 
The concentration of dust in the dead zone assists the warming of the dead zone and the burst triggering by means of the increased optical depth.

-- The burst occurrence frequency depends on the disk evolutionary stage. Initially, the bursts are difficult to resolve because the innermost disk regions switch frequently from the MRI-dead to the MRI-active state and back. In this very early stage characterized by strong disk mass-loading from the envelope, the accretion is highly variable on 1--2 orders of magnitude. This stage is also deeply embedded and therefore may be difficult to probe observationally. In the subsequent evolution, when the mass infall rate onto the disk begins to decline, individual accretion bursts begin to occur and their frequency declines with time. The dependence of the burst frequency on the evolutionary epoch is likely caused by the disk transition from a highly gravitationally unstable state (in which both gravitational and viscous torques provide fast recharge times for the bursts) to a gravitationally stable state, in which mass is inward-transported only by viscous torques, taking longer recharging times (see also \citet{Zhu2009a}.

-- The burst phenomenon depends on the initial prestellar core mass. It is strongest in more massive cores and virtually diminishes for core masses below $0.21~M_\odot$ or the final stellar mass below $M_{\rm \ast,fin}=0.17~M_\odot$. Objects with these core masses form disks of such a low mass and temperature that the thermal ionization of alkaline metals can hardly be achieved, at least in the  considered disk regions beyond 0.5~au.

-- The burst phenomenon was confirmed to occur for a wide range of the mass-to-flux ratios $\lambda$=2--10. The burst occurrence frequency is weakly affected by the variations of $\lambda$ in the considered limits.

The MRI-triggered bursts and dust rings in the inner disk regions are intrinsically linked with each other in our model. Both phenomena  are a manifestation of the dead zone that forms in the innermost disk regions. However, we should warn against over-interpreting this causal link. Accretion bursts in general and dust rings can be caused by phenomena other than the dead zone. For instance, accretion bursts can be triggered by in-falling gaseous clumps in gravitationally unstable disks \citep{VB2005,VB2015}, planet-disk interaction \citep{Lodato2004,Nayakshin2012},  or external perturbations \citep{Bonnell1992,Pfalzner2008,Forgan2010}. Dust rings can be created, in particular, by planets \citep{Rice2006,Dong2015} and snow lines \citep{Zhang2015}. Further studies of the MRI-triggered burst phenomenon are needed to better understand the possible interplay between the dead zone, bursts, and dust rings in protoplanetary disks.

{Our model includes many physical effects with different levels of approximations, which may affect our conclusions. The neglect of non-ideal MHD effects can influence the distribution of magnetic fields in the disk. More accurate dust opacities that take the dust growth into account are needed to better describe the thermal balance in the disk. The ionization fraction calculations can be improved by including compact chemical reaction networks \citep[see, e.g.,][]{Dapp2012} and thermal sublimation of dust grains. Finally, the dust evolution model would benefit from extension to a full multi-bin version and inclusion of dust ice coating into the calculations of dust fragmentation velocity would allow a more accurate calculation of the maximum size of dust grains \citep{Molyarova2020}.   
}


\section*{Acknowledgements}
We thank the anonymous referee for useful comments that helped to improve the manuscript.
This paper is supported by the Austrian Science Fund (FWF) under research grant I2549-N27 and Swiss National Science Foundation (SNSF) (project number 200021L\_163172). The work of Sergey Khaibrakhmanov in Sections~\ref{magfield} and~\ref{sec:adaptive_alpha} is supported by the Large Scientific Project of the Russian Ministry of Science and Higher
Education `Theoretical and experimental studies of the formation and evolution of extrasolar planetary systems and characteristics of exoplanets' (project No.
075-15-2020-780). SB was supported by a Discovery Grant from the Natural Sciences and Engineering Research Council (NSERC) of Canada.
The simulations were performed on
the Vienna Scientific Cluster (VSC-3 and VSC-4) and on the Shared Hierarchical Academic Research Computing Network (SHARCNET).



\end{document}